\documentclass{aa}
\usepackage{graphicx}
\usepackage[varg]{txfonts}
\usepackage{natbib}
\usepackage{varioref}
\usepackage{colortbl}
\bibpunct{(}{)}{;}{a}{}{,} 
\usepackage{hyperref}
\usepackage{siunitx}
\usepackage[all]{hypcap}

\begin{document}

\title{J-PLUS: On the identification of new cluster members in the double galaxy cluster A2589 \& A2593 using PDFs.}
\author{A. Molino\inst{1,2}\fnmsep\thanks{\email{albertomolino.work@gmail.com}} \and
       M. V. Costa-Duarte\inst{1,3} \and C. Mendes de Oliveira\inst{1} \and A. J. Cenarro\inst{4} \and G. B. Lima Neto\inst{1} \and E. S. Cypriano\inst{1} \and L. Sodr\'e Jr\inst{1} \and P. Coelho\inst{1} \and M. Chow-Mart\'inez\inst{5,6} \and R. Monteiro-Oliveira \inst{7,1} \and L. Sampedro\inst{1,2} \and D. Cristobal-Hornillos\inst{4} \and J. Varela\inst{4} \and A. Ederoclite\inst{4} \and A. L. Chies-Santos\inst{7} \and W. Schoenell\inst{1} \and T. Ribeiro\inst{8} \and A. Mar\'in-Franch\inst{4} \and C. L\'opez-Sanjuan\inst{4} \and  J. D. Hern\'andez-Fern\'andez\inst{1} \and A. Cortesi\inst{1} \and H. V\'azquez Rami\'o\inst{4} \and W. Santos Jr\inst{1} \and N. Cibirka\inst{1} \and P. Novais\inst{1} \and E. Pereira\inst{1} \and J. A. Hern\'andez-Jimenez\inst{1} \and Y. Jimenez-Teja$^{9}$ \and M. Moles\inst{4} \and N. Ben\'itez\inst{2} \and R. Dupke\inst{9}.}

\titlerunning{On the identification of new cluster members using PDFs.}
\authorrunning{Molino et al.}
\institute{%
    Universidade de S\~ao Paulo, Instituto de Astronomia, Geof\'isica e Ci\^encias Atmosf\'ericas, Rua do Mat\~ao 1226, 05508-090, S\~ao Paulo, Brazil.
  \and
    Instituto de Astrof\'isica de Andaluc\'ia. IAA-CSIC. Glorieta de la astronom\'ia S/N. 18008, Granada, Spain.
  \and  
    Leiden Observatory, Leiden University, Niels Bohrweg 2, 2333 CA Leiden, The Netherlands
  \and
     Centro de Estudios de F\'isica del Cosmos de Arag\'on, Plaza San Juan 1, 44001 Teruel, Spain. 
  \and
    Departamento de Astronom\'ia, DCNE, Universidad de Guanajuato, Apdo. Postal 144, CP 36000, Guanajuato, Mexico.
  \and
  Instituto de Geolog\'ia y Geof\'isica - Centro de Investigaciones Geocient\'ificas/Universidad Nacional Aut\'onoma de Nicaragua, Managua, Rotonda Universitaria 200 metros al Este, Managua, Nicaragua.
   \and
   Departamento de Astronom\'ia, Instituto de F\'isica, Universidade Federal do Rio Grande do Sul, 15051 Porto Alegre, R.S, Brazil.
  \and
    Departamento de F\'isica, Universidade Federal de Sergipe, Av. Marechal Rondon s/n, 49100-000 São Crist\'ovao, SE, Brazil. 
   \and
   Observat\'orio Nacional, COAA, Rua General Jos\'e Cristino 77, 20921-400, Rio de Janeiro, Brazil.    
}

\date{Accepted.}
\date{Accepted to A\&A on December 2017}

\abstract{
 \textbf{\textit{Aims}}. We aim to use multiband imaging from the Phase-3 Verification Data of the J-PLUS survey to derive accurate photometric redshifts (photo-z) and look for potential new members in the surroundings of the nearby galaxy clusters A2589 ($z$=0.0414) \& A2593 ($z$=0.0440), using redshift probability distribution functions (PDF). The ultimate goal is to demonstrate the usefulness of a 12-band filter system in the study of largescale structure in the local universe. \\
   \textbf{\textit{Methods}}. We present an optimized pipeline for the estimation of photometric redshifts in clusters of galaxies. This pipeline includes a PSF-corrected photometry, specific photometric apertures capable of enhancing the integrated signal in the bluest filters, a careful recalibration of the photometric uncertainties and accurate upper-limit estimations for faint detections. To foresee the expected precision of our photo-z beyond the spectroscopic sample, we designed a set of simulations in which real cluster galaxies are modeled and reinjected inside the images at different signal-to-noise ratio (S/N) levels, recomputing their photometry and photo-z estimates.\\
    \textbf{\textit{Results}}. We tested our photo-z pipeline with a sample of 296 spectroscopically confirmed cluster members with an averaged magnitude of $<r>=16.6$ and redshift $<z>$=0.041. The combination of seven narrow and five broadband filters with a typical photometric-depth of $r\sim21.5$ provides $\delta_{z}$/(1+$z$)=0.01 photo-z estimates. A precision of $\delta_{z}$/(1+$z$)=0.005 is obtained for the 177 galaxies brighter than magnitude $r<$17. Based on simulations, a $\delta_{z}$/(1+$z$)=0.02 and $\delta_{z}$/(1+$z$)=0.03 is expected at magnitudes $<r>=18$ and $<r>=22$, respectively. Complementarily, we used SDSS/DR12 data to derive photo-z estimates for the same galaxy sample. This exercise  demonstrates that the wavelength-resolution of the J-PLUS data can double the precision achieved by SDSS data for galaxies with a high S/N. Based on the Bayesian membership analysis carried out in this work, we find as much as 170 new candidates across the entire field ($\sim$5deg$^2$). The spatial distribution of these galaxies may suggest an overlap between the systems with no evidence of a clear filamentary structure connecting the clusters. This result is supported by X-ray Rosat All-Sky Survey observations suggesting that a hypothetical filament may have low density contrast on diffuse warm gas. \\ 
   \textbf{\textit{Conclusions}}. We prove that the addition of the seven narrow-band filters make the J-PLUS data deeper in terms of photo-z-depth than other surveys of a similar photometric-depth but using only five broadbands. These preliminary results show the potential of J-PLUS data to revisit membership of groups and clusters from nearby galaxies, important for the determination of luminosity and mass functions and environmental studies at the intermediate and low-mass regime. 
  }
  
\keywords{cosmology: large-scale structure of Universe - galaxies: distances and redshifts - galaxies: photometry - galaxies: clusters: general - surveys}

\maketitle

\section{Introduction}
\label{introduction}

Modern astronomy has entered a new era of massive data acquisition. The current and new generation of photometric redshift surveys such as the Sloan Digital Sky Survey (SDSS; \citealt{2000AJ....120.1579Y}), Pan-STARRS (\citealt{2002SPIE.4836..154K}), the Dark Energy Survey (DES; \citealt{2016MNRAS.460.1270D}), the Large Synoptic Survey Telescope (LSST; \citealt{2008arXiv0805.2366I}), the Baryon Oscillation Spectroscopic survey (BOSS; \citealt{2009astro2010S.314S}), EUCLID (\citealt{2010arXiv1001.0061R}), the Dark Energy Spectroscopic Instrument (DESI; \citealt{2013arXiv1308.0847L}) or the Javalambre-Physics of the Accelerated Universe Astronomical survey (J-PAS; \citealt{2009ApJ...692L...5B}, \citealt{2014arXiv1403.5237}) among others, will provide either multicolor or spectral information for millions of galaxies, enabling precise cosmological studies at different cosmic epochs. The combination of deep observations over large angular scales will allow these surveys to reach much larger cosmic volumes, dramatically reducing previous systematic biases due to cosmic variances in specific lines-of-sight (\citealt{2015A&A...582A..16L} and references therein).

\vspace{0.2cm}

In this context, photo-zs have become an essential tool in modern astronomy since they represent a quick and relative inexpensive way of retrieving redshift estimates for a large amount of galaxies in a reasonable amount of observational time. In the last few decades, photometric redshift surveys have mainly been undertaken following two pathways: higher wavelength resolution and moderate depth using medium-to-narrow filters versus deeper observations with a poor resolution using standard broadband filters. The strong dependency between the wavelength resolution (number and type of passbands) and the achievable precision of photo-z estimates (\citealt{1994MNRAS.267..911H}, \citealt{1998ApJS..115...35H}, \citealt{2001A&A...365..660W}, \citealt{2009ApJ...691..241B}) has inspired the design of a whole generation of medium-to-narrow multiband photometric redshift surveys such as the Classifying Object by Medium-Band Observations-17 survey (COMBO-17; \citealt{2003A&A...401...73W}), the MUltiwavelength Survey by Yale-Chile (MUSYC; \citealt{2006ApJS..162....1G}), the Advance Large Homogeneous Medium Band Redshift Astronomical survey (ALHAMBRA; \citealt{Molesetal2008}), the Cluster Lensing and Supernovae with Hubble survey (CLASH; \citealt{2012ApJS..199...25P}), the Survey for High-z Absorption Red and Dead Sources (SHARDS; \citealt{2013ApJ...762...46P}) among others, reaching as accurate photo-z estimates as $\delta_{z}$/(1+$z$)<0.01 for high signal-to-noise ratio (S/N) galaxies. Meanwhile, very deep broadband photometric observations such as the Hubble Deep Field (HDF; \citealt{1995AAS...187.0901F}), the Hubble Ultra-Deep Field (HUDF; \citealt{2006AJ....132.1729B}), the Cosmic Assembly Near-infrared Deep Extragalactic Legacy Survey (CANDELS; \citealt{2011ApJS..197...35G}), the Hubble Extreme Deep Field (XDF; \citealt{2013ApJS..209....6I}), the Hubble Frontiers Field program (HFF; Lotz et al., in prep.) or the REionization LensIng Cluster Survey (RELICS; Coe et al., in prep.) among others, even with a limited photo-z accuracy of $\delta_{z}$/(1+$z$)$>$0.05, have extended our current knowledge on the formation and evolution of galaxies all the way back to a $z$>10-12, posing a new tension between theory and observations regarding when and at which rate the first generation of galaxies was formed and how it contributed to the cosmic reionization of the Universe (\citealt{2013ApJ...768...71R}). 
 
\vspace{0.1cm}

After the pioneering works of \cite{1958ApJS....3..211A}, \cite{1961cgcg.book.....Z}, \cite{1969AJ.....74..804K}, \cite{1970AJ.....75..237S}, \cite{1974AJ.....79.1356C}, \cite{1977ApJS...35..209D} or \cite{1978MNRAS.182..241B}, visually searching for rich clusters of galaxies in the local Universe, the detection and cataloging of groups and clusters of galaxies in both hemispheres went through a tremendous development in the subsequent three decades. With the advent of a new technological era, it was possible to replace original photographic plates by scanned plates or CCD imaging. This represented a major change in the study of cluster of galaxies since it made feasible not only to extend searches to a deeper magnitude limit (or a lower-surface brightness) but also to derive more homogeneously selected samples (\citealt{1983ApJ...268..476S}; \citealt{1989ApJS...70....1A}; \citealt{1990ApJ...365...66H}). Likewise, the combination of widefield optical CCD imagers with the advent of large telescopes made possible to pass from discrete programs aiming at observing samples composed by a few thousand systems to global all-sky surveys such as the ESO-Nearby Abell Cluster survey (\citealt{1996A&A...310....8K}), the Las Campanas Catalogue (\citealt{2000ApJS..130..237T}), the Northern Sky Optical Cluster Survey (NoSOCS; \citealt{2003AJ....125.2064G}), the Digitalized Second Palomar Observatory Sky Survey (DPOSS; \citealt{2003AJ....125.2064G}), the 6dF Galaxy Redshift survey (6dFGS; \citealt{jones2004}) or the Sloan Digital Sky Survey (SDSS; \citealt{2011ApJ...736...21S}), among others.   

\vspace{0.2cm} 

Although redshift estimations from galaxy colors are more uncertain than those obtained directly from a spectrum, this situation is certainly being improved. The new generation of multi narrow-band surveys such as COMBO-17 (\citealt{2001A&A...365..660W}), COSMOS-21 (\citealt{2007ApJS..172....9T}), ALHAMBRA (\citealt{Molesetal2008}), COSMOS-30 (\citealt{2009ApJ...690.1236I}), J-PLUS (\citealt{Cenarro18}), S-PLUS (Mendes de Oliveira et al., in prep.) or J-PAS (\citealt{2014arXiv1403.5237}) can provide ``very-low resolution spectra" with which retrieve as accurate photo-z estimates as $\delta_{z}$/(1+$z$)=0.005 for high S/N galaxies (see Section \ref{accspeczdata} of this paper). As demonstrated in this work, although the J-PLUS observations are similar in terms of photometric-depth to those from SDSS, they are indeed deeper in terms of photo-z-depth due to the additional seven narrowbands present in the filter system. The application of photo-z (or color-based) methods for the identification of clusters of galaxies is relatively new (\citealt{1994AJ....108.1476F}; \citealt{1995A&A...297...61B}; \citealt{2000AJ....119...12G}; \citealt{2000AJ....120..540G}; \citealt{2003AJ....125.2064G}; \citealt{2009AJ....137.2981G}; \citealt{2012ApJS..199...34W}; \citealt{2012MNRAS.420.1167A}, \citealt{2014MNRAS.439.1980A}, \citealt{2015MNRAS.452..549A}, among others). However, as emphasized by several authors (\citealt{2000ApJ...536..571B}; \citealt{2002MNRAS.330..889F}; \citealt{2006AJ....132..926C}; \citealt{2008MNRAS.386..781M}; \citealt{2009MNRAS.396.2379C}; \citealt{2009A&A...508.1173P}; \citealt{2009ApJ...700L.174W}; \citealt{2010MNRAS.406..881B}; \citealt{2011ApJ...734...36A}; \citealt{2012ApJS..201...32S}; \citealt{2013MNRAS.432.1483C}; \citealt{2014MNRAS.441.2891M}; \citealt{2014MNRAS}; \citealt{2015A&A...582A..16L}; \citealt{2015A&A...576A..25V}; \citealt{2016arXiv161109231L}, among others), if photo-z are treated as PDFs rather than simple point-estimates, robust and systematic analysis can be carried out in many extra-galactic and cosmological studies down to fairly deep magnitudes, retrieving similar results to those obtained when using spectroscopic redshift information.      
 
\vspace{0.2cm}

This work represents an effort on that direction. The methodology discussed here does not aim to be another cluster-finder algorithm but rather to show the potential of the J-PLUS data to find new and/or fainter members in previously known groups or galaxy cluster systems, based on the analysis of PDFs derived from accurate photo-z estimations. Taking into account that the new generation of photometric redshift surveys will surpass the photometric-depth of current spectroscopic redshift surveys such as SDSS ($m_{rSDSS}<$17.77, \citealt{2015ApJS..219...12A}), they will allow detection and characterization of the faintest populations in the nearby Universe, shedding new light to the underlying distribution of the dark matter. In the case of clusters of galaxies, these data will allow us to carry out systematic studies of the faintest populations ($r>18$) composing these systems, leading to a better determination of cluster memberships, a more accurate derivation of luminosity functions and a better overall understanding on the formation and evolution of clusters of galaxies. Therefore, this paper aims to highlight the potential of the J-PLUS survey to revisit previous knowledge of the largescale structure in the nearby Universe.

\vspace{0.2cm}

This paper is organized as follows. In Section \ref{data}, we present the data used in this work, a combination of both optical multiband photometric data and an optical spectroscopic sample of cluster galaxies. In Section \ref{photometry}, we present the optimized pipeline adopted in this work to derive an accurate PSF-corrected aperture-matched photometry among passbands for galaxies in clusters along with an empirical characterization of the depth of our images. In Section \ref{stargalaxy} we briefly introduce the methodology adopted here for the well-known star-galaxy separation problem. Section \ref{photozs} is devoted to the description and quantification of the photo-z estimates obtained on real data and on simulations. A systematic comparison between SDSS and J-PLUS is done to probe the benefit of increasing the number of passbands. Once the photo-z catalog has been presented and tested, in Section \ref{ClusterMembers} we present the methodology proposed for the identification of new cluster members along with several tests used to quantify the reliability of this technique. Finally, in Section \ref{discussion}, we discuss the results obtained, highlighting the possibility of extending this sort of analysis to other nearby galaxy clusters when the complete J-PLUS data becomes available. Unless specified otherwise, all magnitudes here are presented in the AB system. Throughout this work, we have adopted the cosmological model provided by the Planck collaboration (2014) with parameters $h_{0}$ = 0.7 km$s^{-1}$ Mpc$^{-1}$ and ($\Omega_{M}$, $\Omega_{\Lambda}$, $\Omega_{K}$) = (0.31, 0.69, 0.00).

\section{Data}
\label{data}

\subsection{Observations}
\label{observations}

\begin{figure}
\centering
\includegraphics[width=9.cm]{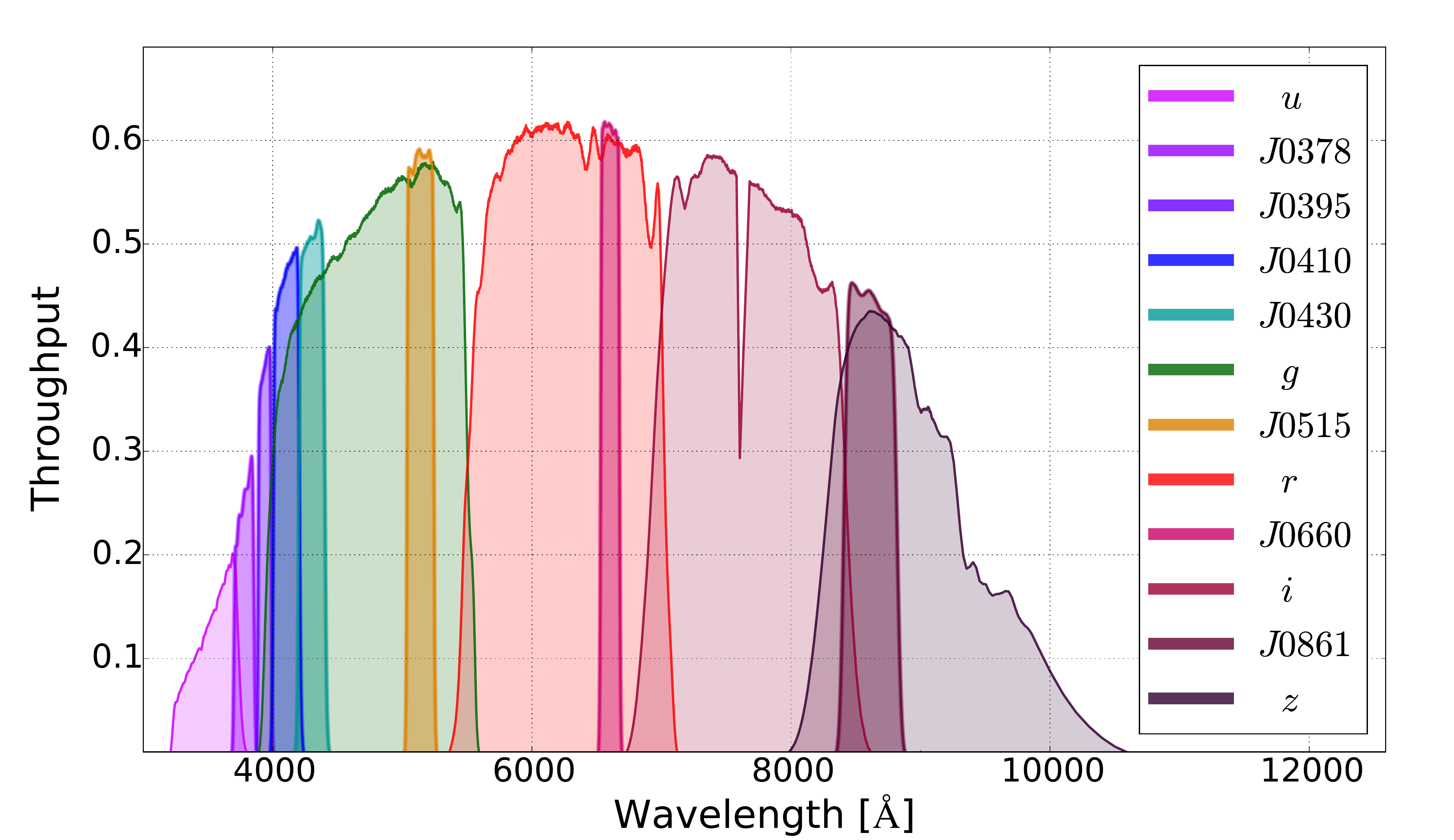}
\caption[the J-PLUS filter system]{Final throughput of the J-PLUS filter system. It incorporates five broad ($u$, $g$, $r$ $i$ \& $z$) \& seven narrow ($J0378$, $J0395$, $J0410$, $J0430$, $J0515$, $J0660,$ \& $J0861$) bands covering the entire optical range.}
\label{filtersystem}
\end{figure}

All observations of Abell2589 \& Abell2593 were collected during the nights of February 25 and 26, 2016 as part of the Phase-3 Verification data of the Javalambre-Photometric Local Universe Survey (J-PLUS\footnote{https://www.j-plus.es}; \citealt{Cenarro18}). This survey uses a 0.83 meter telescope (JAST/T80) at the Observatorio Astrof\'isico de Javalambre (OAJ)\footnote{http://oajweb.cefca.es} in Spain, and an optical imager (T80Cam) composed by a 9216$\times$9232 pixel array, with a 0.55"/pixel-scale and a 1.4$\times$1.4 deg$^{2}$ FoV. The J-PLUS survey uses a photometric system composed of five broadband ($u$, $g$, $r$, $i,$ \& $z$) \& seven narrowband ($J0378$, $J0395$, $J0410$, $J0430$, $J0515$, $J0660$ \& $J0861$) filters covering the whole optical range. Figure $\ref{filtersystem}$ shows the final throughput of the J-PLUS filter system. As throughly described in Section 6 of \cite{2014arXiv1403.5237}, this is an optimized filter system for the identification of stars and the calibration of the J-PAS survey. We refer the reader to Table $\ref{tablefilter}$ on this paper or to \citealt{Cenarro18} for more details about this filter system. 

\vspace{0.2cm}

\begin{figure*}[tbp]
\centering
\includegraphics[width=18.cm]{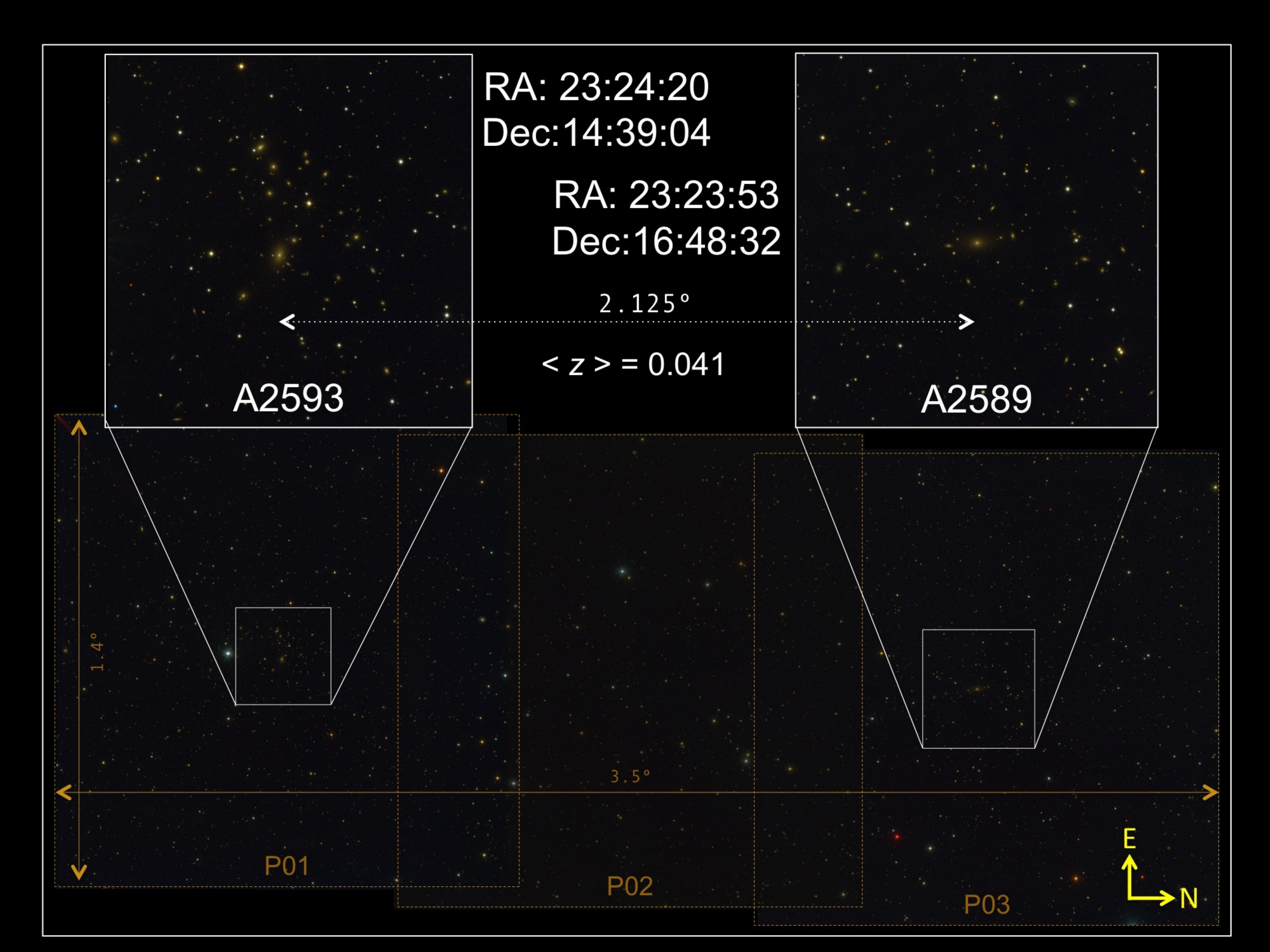}
\caption[The A2589 \& A2593 galaxy clusters]{A2589 \& A2593 galaxy clusters. The figure shows the final layout for the observations composed by three overlapping J-PLUS pointings, covering a total area of 3.5 x 1.4 deg$^{2}$. Central coordinates for Pointing-1 and Pointing-3 (centered at the cluster BCGs) along with the physical separation among the clusters are indicated in the upper central region of the figure. This RGB color image has been created with the \textit{TRILOGY} software (http://www.stsci.edu/$^{\sim}$dcoe/trilogy).}
\label{mosaico}
\end{figure*}

To guarantee a relative photometric calibration across the entire field (see Section $\ref{sdss2jplus}$), observations were divided in three pointings with a small ($\sim$0.3 deg$^{2}$) overlap. The central coordinates of the three pointings were ($\alpha$, $\delta$) = (351.0025$^{o}$, +16.8068$^{o}$) for Pointing-1, ($\alpha$, $\delta$) = (351.0315$^{o}$, +15.7560$^{o}$) for Pointing-2 and ($\alpha$, $\delta$) = (351.0944$^{o}$, +14.6469$^{o}$) for Pointing-3. Figure $\ref{mosaico}$ shows a color image with the final observational layout encompassing the three pointings. 

\vspace{0.2cm}

\begin{figure}
\centering
\includegraphics[width=9.3cm]{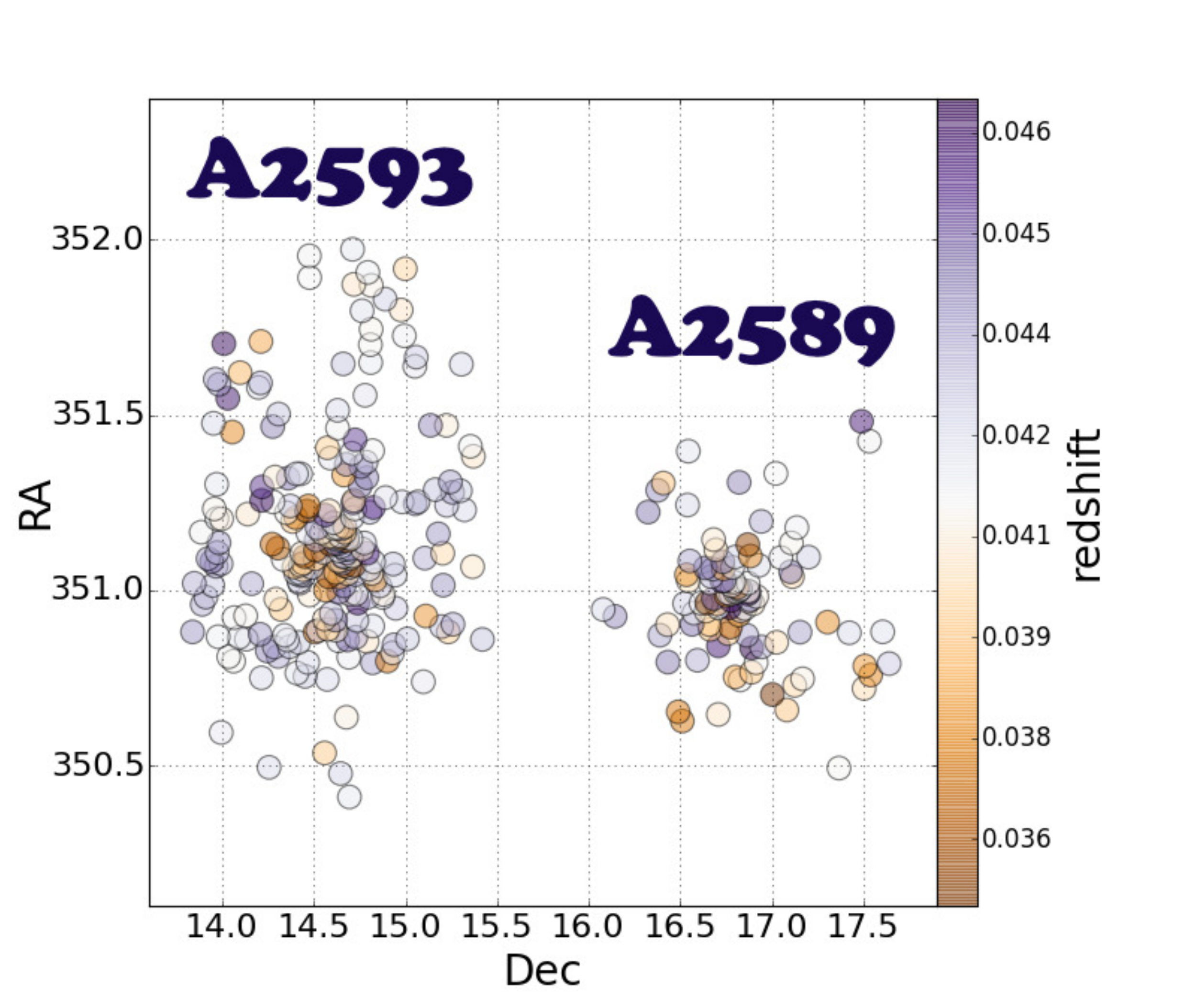}
\caption[spectroscopic sample.]{Spectroscopic redshift sample. The figure shows the spectroscopic redshift sample used in this work, composed by $\sim$300 confirmed galaxy cluster members: 97 from A2589 and 222 from A2593. The sample was used for the characterization of the J-PLUS photo-z on cluster galaxies (Section $\ref{photozquality}$).}
\label{speczsample1}
\end{figure}

Taking advantage of the fact that data from the SDSS/DR12 (\citealt{2015ApJS..219...12A}) are available for all the three pointings and that both surveys have a similar photometric depth, we were able to find as much as 300k detections in common between both surveys. As explained in Section \ref{photometry} this sample served to quantify the level of agreement between datasets and to guarantee an absolute photometric calibration across the entire field. In order to characterize the robustness of our photo-z estimations and the performance of the cluster membership method presented below, we select a spectroscopic sample of $\sim$300 galaxies (as shown in Figure $\ref{speczsample1}$), with 97 galaxies from A2589 and 222 galaxies from A2593, and an average magnitude $<r>\sim$16.6. 
This control sample was extracted from the Andernach's compilation of radial velocities for galaxies that are probable members of Abell/ACO clusters (presented in \citealt{ander2005})\footnote{The compilation consists of a collection of individual radial velocities for galaxies inside or close to the Abell radius for about 3\,930 clusters (75\% of the total). The latest version of the compilation (2015), which was used in this work, provides redshifts for about 130\,000 individual cluster galaxies. We refer the interested reader to the Appendix section for additional details.}. The radial velocities have been taken from sources of the published (and, in few cases, unpublished) literature. The published data are from both individual published data as well as large-scale multiobject spectroscopic surveys such as the Two-Degree Field Galaxy Redshift Survey (2dFGRS; \citealt{colles2001}), the Six-Degree Field Galaxy Redshift Survey (6dFGS; \citealt{jones2004}) and the Sloan Digital Sky Survey \citep{abazajian2009}. In addition to the projected distance of the galaxy to the Abell cluster center, a gap criteron ($\pm 1\,500$ km s$^{-1}$) is used to separate one concentration of another in redshift space. Galaxies exceeding any concentration in the line-of-sight are maintained in the compilation but flagged as nonmembers. These galaxies were not considered in this work.

Finally, in Section $\ref{ClusterMembers}$, we make use of the photo-z catalog presented in \citealt{ShunFang2011} (hereafter SF11), to compare (for the specific case of A2589) the number of new members found in both studies.

\subsection{Data processing and calibration.}
\label{pipeline}

The storage, processing and calibration of the J-PLUS data was carried out using the automatized Jype pipeline developed and implemented at the Centro de Estudios del Cosmos de Arag\'on (CEFCA), Spain. Some technical details are presented in \cite{2014SPIE.9152E..0OC} and a more detailed description of the process will be presented in \citealt{Cenarro18}. Here we briefly summarize the key steps to convert raw data products to science images. In order to remove the instrumental signatures from images overscan subtractions, flat-field corrections, bad pixel and/or cosmic-ray rejections and (if needed) fringe corrections were applied to the science images. Astrometric calibration of images was computed using SCAMP (\citealt{2006ASPC..351..112B}) reaching an accuracy at the level of a fraction of a PSF ($\sim$55mas). The relative photometry among the images is computed in order to be scaled to the same photometric reference. Image stacking was done using SWARP (\citealt{2002ASPC..281..228B}). An initial photometric zero-point calibration is performed for final reduced and co-added science images, using SDSS/DR12 aperture photometry of stars in the field (for the broad-bands), and the synthetic photometry computed from the  SDSS spectra (for the narrow-bands). Final zero-point refinements were done using photometric redshift estimations (see Section \ref{photozpcal}). 

\vspace{0.2cm}

\begin{table}
\caption{{\small Summary of the multiwavelength filter set for J-PLUS. The FWHM, the exposure time and the limiting magnitude (measured on 3" diameter aperture) correspond to the average value among the three pointings. Since this is Science Verification Data, the photometric depth presented in this table may differ from the values adopted for the main survey.}}
\begin{center}
\label{tablefilter}
\begin{tabular}{|l|c|c|c|c|c|c|c|c|c|c|c|c|c|c|c|c|}
\hline
Filter & $\lambda_{eff}$ & $\Delta \lambda$  & Frames & Total & Seeing & $m_{lim}^{(3")}$ \\
\hline
name & [nm] & [nm] & [\#] & [sec] & ["] & (5-$\sigma$) \\
\hline
\,\,\,\, $u$    &  350  &   38    &  3    &  588  & 1.42 & 21.28 \\
$J0378$    &  378  &   13    &  3    &  540  & 1.52 & 20.69 \\
$J0395$    &  395  &   7     &  3    &  278  & 1.48 & 20.39 \\
$J0410$    &  410  &   18    &  3    &  116  & 1.63 & 20.18 \\
$J0430$    &  430  &   18    &  3    &  111  & 1.80 & 20.15 \\ 
\,\,\,\, $g$    &  483  &   135   &  3    &  39   & 1.60 & 21.01 \\
$J0515$    &  515  &   18    &  3    &  123  & 1.42 & 20.16  \\
\,\,\,\, $r$    &  631  &   135   &  3    &  60   & 1.38 & 21.36 \\
$J0660$    &  660  &   11    &  3    &  809  & 1.46 & 21.81 \\
\,\,\,\, $i$    &  778  &   139   &  3    &  78   & 1.30 & 21.28  \\
$J0861$    &  861  &   36    &  3    &  180  & 1.37 & 20.45  \\
\,\,\,\, $z$    &  920  &   153   &  3    &  108  & 1.37 & 20.43 \\
\hline
\end{tabular}
\end{center}
\end{table}   
 
\section{Accurate multiband photometry.}
\label{photometry}

This section is devoted to the explanation of how a multiband aperture-matched PSF-corrected photometry has been performed on this dataset. In Section $\ref{colorpro}$ we initially present the methodology applied to deal with images with different PSF conditions and provide accurate colors for photo-z estimations. In Section $\ref{sdss2jplus}$, we explain how we have derived a relative photometric calibration among the three pointings. In Section $\ref{photodepth}$, we characterize the photometric-depth of our images. 

\subsection{PSF-corrected aperture-matched photometry.} 
\label{colorpro}

In recent decades, many photometric surveys have progressively make use of a larger number of passbands, in many cases replacing standard broadband for narrowband filters, to increase the sensitivity in the detection of emission-lines with moderate equivalent widths. Enhancing the wavelength resolution of observations serves not only to better reconstruct the spectral energy distribution (SED) of astronomical sources, certainly improving their type classification, but also to boost the quality of redshift estimation of galaxies from photometric colors. While traditional four- to five-passband surveys provide redshift estimates with typical errors $\delta_{z}$/(1+$z$)$\sim$3-4\% (such as SDSS (\citealt{2003AJ....125..580C}, \citealt{2012ApJ...747...59R}) or DES (\citealt{2014MNRAS.445.1482S})), 10- to 40-passband survey can reduce this scatter to a level of $\sim$1-2\% (CFHTLS (\citealt{2006A&A...457..841I}), COMBO-17 (\citealt{2008A&A...492..933W}), COSMOS (\citealt{2009ApJ...690.1236I}), MUSYC (\citealt{2010ApJS..189...270C}), ALHAMBRA (\citealt{2014MNRAS.441.2891M}; \citealt{2017MNRAS.464.4331N}), COSMOS2015 (\citealt{2016ApJS..224...24L}), CLASH (\citealt{2017MNRAS.470...95M})), among others. As demonstrated in \cite{2009ApJ...691..241B}, the incoming new generation of 50-60 narrowband photometric surveys will reach as accurate redshift estimations as $\delta_{z}$/(1+$z$)$\sim$0.3\% for million galaxies (J-PAS (\citealt{2014arXiv1403.5237}), PAU (\citealt{2014MNRAS.442...92M})). We also refer the reader to figure B1 in \cite{2014MNRAS.441.2891M} for an illustration of this tendency in the evolution of the number of filters.   

\vspace{0.2cm}
 
However, this refinement or improvement in the photo-z estimations due to the systematic increasing in the wavelength resolution has also brought new technical challenges to the field. Adding more passbands to a survey represents a multiplicative factor in the observational time since specific sky regions need to be imaged several times until all filters are observed and they reach the final desired depth. For large and deep programs, the completion of the data, therefore, may only be accomplished after long periods of time (from days to years). This time-lapse effect may lead to a large variation in the quality of the observations such as seeing, sky-brightness, airmass or extinction. Sources of inhomogeneity that, if not properly taken into account, may worsen the performance of photo-z estimates. 

\vspace{0.2cm}

To retrieve the best redshift estimations possible for a given dataset, an important effort needs to be devoted to the homogenization and calibration of the photometry. In order to derive an accurate photometry across the different bands, we have followed a similar approach as that presented in \cite{2014MNRAS.441.2891M} for the ALHAMBRA survey, to correct differences in the seeing conditions over different images. Basically, a PSF-corrected photometry has been performed using an updated version of the ColorPro software (\citealt{2006AJ....132..926C}; \citealt{2014MNRAS.441.2891M}). This software compensates artificial differences in the magnitudes of galaxies when using a fixed photometric aperture across a set of images with a nonhomogeneous PSF. It also makes it possible to preserve the best quality images without the need of degrading the whole system to the worst condition - in other words, preserving all the information from the best quality images.    

\vspace{0.2cm}

\begin{figure}
\centering
\includegraphics[width=9.cm]{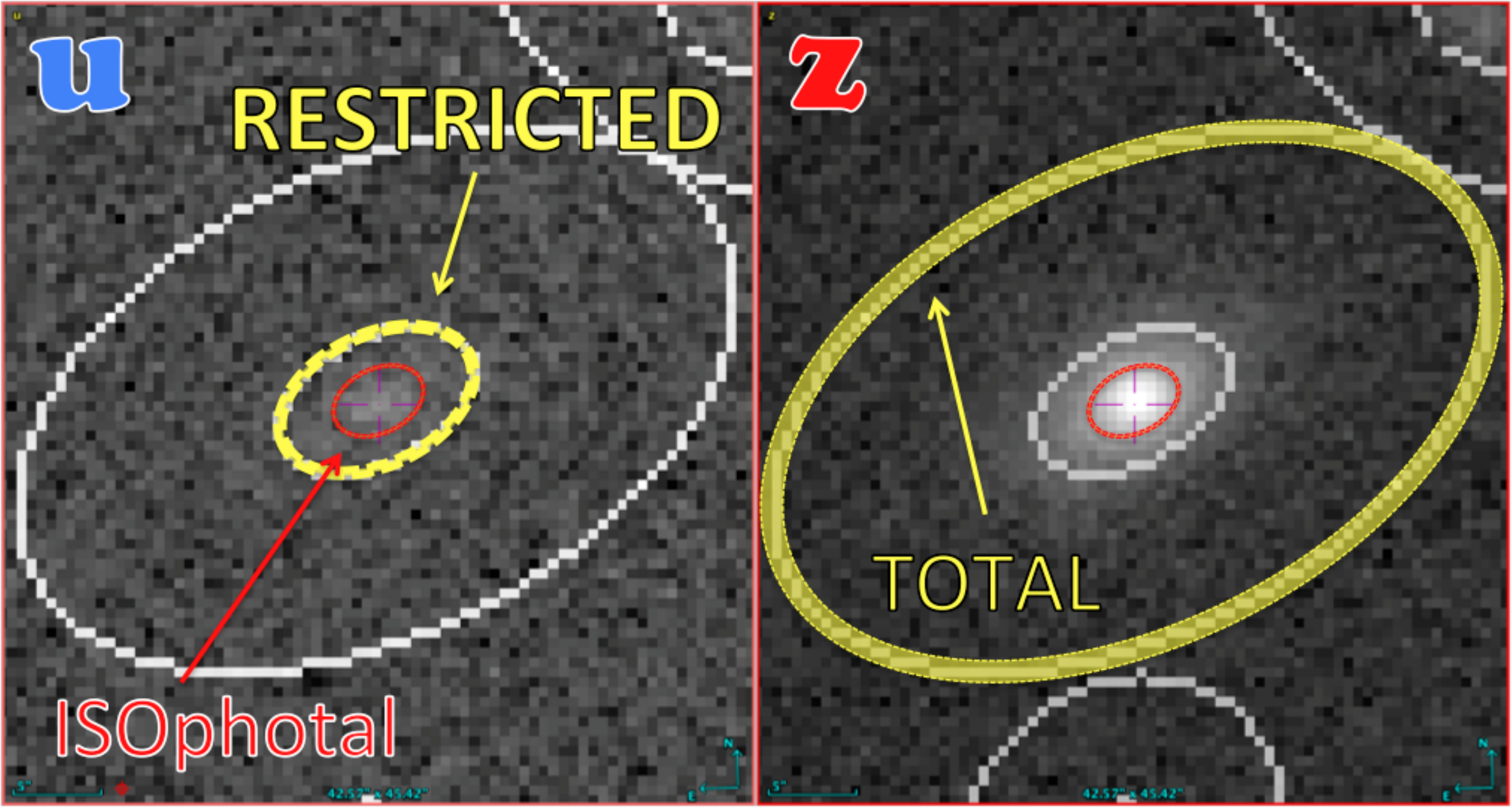}
\caption{Photometric apertures for cluster galaxies. The figure illustrates, for a typical cluster galaxy, how standard apertures defined on deep reddish images (right panel) lead to an inefficient photometry on the bluest filters (left panel). In this work we adopt total ``restricted" apertures integrating most of the light from the galaxies while keeping a higher S/N. }
\label{totalrestricted}
\end{figure}

Keeping in mind that the scope of this paper is the identification of potential new cluster members (CMs) inside and in between the galaxy clusters A2589 \& A2593, we created a deeper detection image as a combination of the $r$, $i,$ and $z$ bands. These images served to increase the detectability of sources as well as to improve the photometric apertures definition due to the enhanced S/N of faint sources. However, as demonstrated in \cite{2017MNRAS.470...95M}, for a sample of 25 massive galaxy clusters from the CLASH survey (\citealt{2012ApJS..199...25P}), defining photometric apertures for early-type galaxies based on deep and reddish detection-images led to an under-performance of photo-z estimations due to an artificial deterioration of the S/N in the bluest filters. As discussed in the aforementioned paper, it is possible to circumvent the effect by adopting an optimal photometric aperture definition for early-type galaxies. Briefly, total ``restricted" apertures are created forcing \texttt{SExtractor} (\citealt{1996A&AS..117..393B}) to define total (AUTO) magnitudes but with the smallest possible radius, integrating most of the light from the galaxies while keeping a higher S/N than the standard \texttt{SExtractor} AUTO magnitudes in the shortest wavelengths. Although these restricted apertures may miss a small fraction of the light from the outskirts of the galaxies, they provide much more accurate colors, improving photo-z estimates. Figure $\ref{totalrestricted}$ illustrates such photometric apertures.

\subsection{Photometric cross-checks with SDSS.}
\label{sdss2jplus}

\begin{figure}
\centering
\includegraphics[width=9.cm]{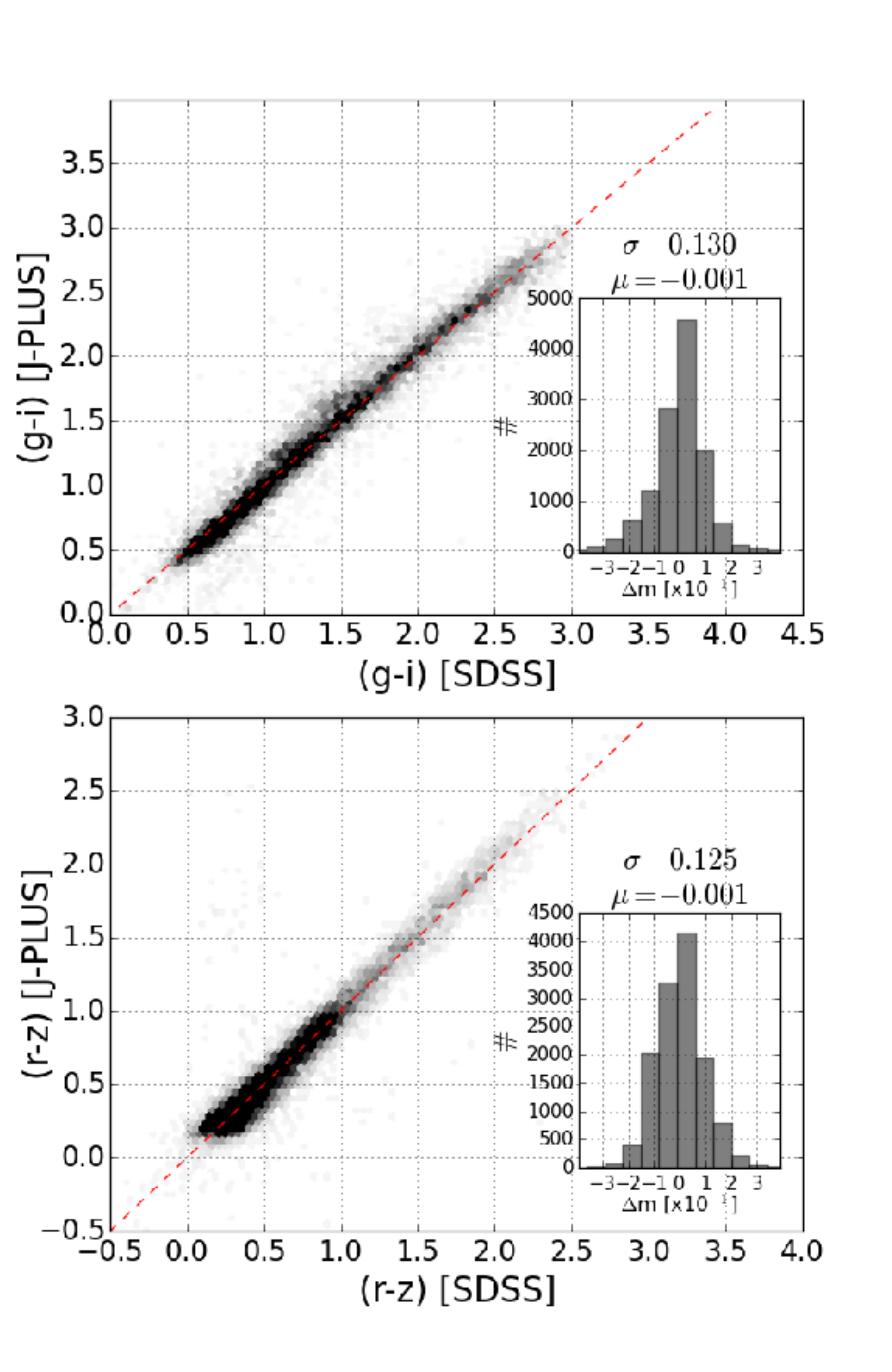}
\caption{Comparison between SDSS/DR12 and J-PLUS data. Taking advantage of the overlap between the two surveys, we selected a sample of $\sim$ 300k common detections to quantify the level of agreement between the two data sets. Colors rather than magnitudes were used since they are less sensitive to different aperture definitions. An intrinsic dispersion of $\sim$10\% ($\sigma=0.13$) and negligible bias ($\mu$=-0.001) was observed for sources as faint as a magnitude $g_{SDSS}$=21.}
\label{sdssjplusplot}
\end{figure}

As mentioned before, we took advantage of the fact that our three pointings have overlap with SDSS/DR12 to systematically compare our photometry with that from SDSS/DR12, and be able to quantify the level of agreement between both data sets. To do so, we looked for sources with good photometry in the SDSS/DR12 data that were also detected in our catalogs, and found as many as 300k detections in common. As illustrated in Figure \ref{sdssjplusplot}, we preferred to rely on color-color diagrams (($g$-$i$) versus ($r$-$z$)) for the comparison of the two datasets since colors are less sensitive than single magnitudes to the specific definition of the photometric apertures from each survey, therefore making the comparison more reliable. Based on these diagrams, we find that our data are capable of reproducing the SDSS colors with an intrinsic dispersion of $\sim$10\% ($\sigma=0.13$) and negligible bias ($\mu$=-0.001) for galaxies as faint as a magnitude $g_{SDSS}$=21. At brighter magnitudes these differences were considerably reduced to $\sigma=0.03-0.05$, depending on the filter. 

\vspace{0.2cm} 

After confirming the relative calibration of each individual pointing with respect to SDSS, we also studied the consistency of the calibration among the three pointings. We compared the magnitudes of repeated detections within overlapping regions; meaning the same sources, but detected in different frames (see layout displayed in Figure $\ref{mosaico}$ for more details). We observed an average dispersion of $<$3\%, compatible with the expected photometric noise at different S/N levels. This exercise served to prove the absolute calibration of the photometry across the entire field. Finally, considering that our blue broadband filter ($u$) slightly differs from that of SDSS ($u_{SDSS}$), we looked for a possible transformation equation to pass from one to the other, following a double approach. We followed the same methodology as presented in \cite{2014MNRAS.441.2891M} using colors of galaxies to estimate the expected transformation equation between filter systems. Based on this methodology, we found the following transformation:

\begin{eqnarray}
\label{US2UJ}
\centering
\,\,u_{SDSS} = 0.03 + 0.67 \times u+0.28 \times J0378+0.05 \times J0395
.\end{eqnarray}

The transformation presented in Equation \ref{US2UJ} provides accurate SDSS colors with a negligible scatter $<$0.5\%, inferior to the expected photometric noise of images. Although this transformation is robust, it has the limitation of needing the magnitudes in all the three ($u$, $J0378$ \& $J0395$) bands. Unfortunately, this condition is not always reachable at faint magnitudes. Therefore, this approach is only suitable for data taken with high S/Ns. We also estimated the equivalence of both filters ($u_{SDSS}$-$u$), simply comparing the magnitudes from both datasets. For galaxies with magnitudes 14$<g_{SDSS}<$20, we found that the color ($u_{SDSS}$-$u$) is well-represented by a normal distribution with a scatter of $\sigma$=0.6 and a $\mu\sim$0.21, as shown in Equation \ref{US2UJ2},

\begin{eqnarray}
\centering
\label{US2UJ2}
\,\,\,\,\,\,\,\,\,\,\,\,\,\,\,\,\,\,\,\,\,\,\,\,\,\,\,\,\,\,\,\,\,\,\,\,\,\,\,\,\,\,\,\,\,\,\,\,\,\,\,u_{SDSS} = u + 0.21
.\end{eqnarray}

Despite the fact that this definition provides a noisier dispersion for the $u_{SDSS}$ magnitudes than the one presented in Equation \ref{US2UJ}, it does have the advantage of not needing the sources to be detected in all the three narrow-band filters, making possible to extend the comparison to fainter magnitudes. It is beyond the scope of this paper to assess which definition is more appropriate and should be adopted. We leave this decision to the user, based on particular science cases.  

\subsection{Photometric-depth of images.}
\label{photodepth}

Photometric upper-limits represent an estimate of faintest signal detectable by an astronomical image. Depending on the redshift range under study, these estimations may be of paramount importance when computing photo-z estimations since they serve to break unwilling degeneracies in the color-redshift space. In order to compute accurate upper-limits for our photo-z estimates, we 
decided to re-estimate the photometric uncertainties in our  catalogs. To do so, we rely on an empirical approach similar to those followed by \cite{2000AJ....120.2747C}, \cite{2003AJ....125.1107L}, \cite{2006ApJS..162....1G}, \cite{2007AJ....134.1103Q}, \cite{2008ApJ...682..985W} or \cite{2014MNRAS.441.2891M}, characterizing the expected background signal added to our magnitudes as a function of the photometric apertures. Basically, we first compute the \texttt{SExtractor} segmentation-map associated to every detection image to find out which pixels are associated to real detections. Then $\sim$50.000 apertures are thrown over the remaining (blank) area, saving the enclosed signal and the RMS inside it. The procedure is repeated for apertures in the 1-20 pixel radius, correcting appropriately by the effective exposure time of the pixels using the corresponding weight-maps. We used the so-estimated empirical aperture-to-noise relation to correct the \texttt{SExtractor} uncertainties.

\vspace{0.2cm}

Taking advantage of this noise characterization, we derived empirical photometric upper-limit estimations for nondetected sources on a particular band. Whenever a source was nondetected (m$_{i}$=99.0), we replaced its photometric uncertainty (em$_{i}$) by an upper-limit ($mag^{n\sigma}_{upp}$) as requested by the photo-z code used in this work (Section \ref{photozs}). We used the size (area in pixels) of the photometric aperture defined in the detection image to estimate the typical background signal enclosed within such apertures. The signal is converted to magnitudes according to Equation \ref{upperlims1}:    

\vspace{0.1cm}

\begin{equation}
\label{upperlims1}
\,\,\,\,\,\,\,\,\,\,\,\,\,\,\,\,\,\,\,\,\,\,\,\,mag^{n\sigma}_{upp} = {-2.5\,\times\,log(n \times \sigma_{rms})\,+zp_{i}}
\end{equation} where $\sigma_{rms}$ denotes the 1-$\sigma$ estimate from the noise distribution, $n$ the number of sigmas requested for the limiting magnitudes and $zp_{i}$ the photometric zero-point of an image. These limiting magnitudes strongly depend on the noise of images, the adopted aperture to compute the photometry and the significance requested for the detection to be considered real.  

\section{The star-galaxy separation.}
\label{stargalaxy}

The J-PLUS survey aims to image $>$8000 square degrees in the northern hemisphere. Given the wide field-of-view (FoV) and the photometric-depth of images (see Section \ref{observations}), at magnitudes brighter than $r<$18 the observations are completely dominated by Galactic sources. Since this paper aims to flag up potential new cluster galaxies in these fields, it became mandatory to previously remove  potential stars from our samples. In this work we have relied on a star-galaxy classification based on random forest (Breiman+01), combining the J-PLUS multiband photometry with the morphological information from the sources (Costa-Duarte, in preparation). Besides, a large spectroscopic sample of both stars and galaxies from the SDSS \citep{2017arXiv170709322A} and GAMA \citep{2011MNRAS.413..971D} survey were used for validation purposes. The optimization process along with the discussion of its reliability as a function of the number of bands a source was detected and its S/N, will be thoroughly explained in the aforementioned paper (in prep.).

\vspace{0.2cm}

After applying the star-galaxy classification method to the whole catalog, we find (on average) that the Galactic contribution represents up to $\sim$75\% of the detected sources in our images. In other words, only $\sim$25\% of the detected sources are classified as potential galaxies. This result emphasizes the importance of ``decontaminating'' the general catalogs from stellar objects for any extra-galactic analysis based on the J-PLUS data.

\section{Photometric redshifts.}
\label{photozs}
We rely on the Bayesian Photometric Redshifts (\texttt{BPZ}) code (\citealt{2000ApJ...536..571B}) to calculate photo-z for our detections. \texttt{BPZ} is a Bayesian template-fitting code where a likelihood function coming from the comparison between data and models is weighted by an empirical luminosity-based prior. The combination of these two pieces of information provides a whole posterior PDF in the redshift vs spectral-type (template) space (i.e., P($z$,$T$)). Compared to the public version from 2000, the \texttt{BPZ2.0} includes a new library of 11 galaxy templates (five for ellipticals, two for spirals and four for starburst) which include emission lines and dust extinction. The opacity of the intergalactic medium is applied as described in \cite{1995qal..conf..377M}. In addition, this new version of \texttt{BPZ} provides an estimate of both the absolute magnitude and the stellar mass content of galaxies based on the most probable redshift and spectral-type solution. We refer the reader to \cite{2014MNRAS.441.2891M} for more details of the \texttt{BPZ2.0}.

\vspace{0.2cm}

In order to compare the quality of our results with other previous or similar works, we have relied on the normalized median absolute deviation (NMAD). The reason for choosing this indicator is twofold. This indicator manages to get a stable estimate of the spread of the core of the photo-z error distribution without being affected by catastrophic errors (therefore causing the photo-z error distribution to depart from Gaussianity). We also adopted it because it has become a standard parametrization on the photo-z performance for many different groups, making simpler the comparison among works (see Section $\ref{photozquality}$). The NMAD indicator is defined as follows:

\begin{equation}
\,\,\,\,\,\,\,\,\,\,\,\,\,\,\,\,\,\, \sigma_{NMAD} = 1.48 \times median({\left | \delta z - median(\delta z) \right | \over 1+z_{s}})
\label{nmad}
\end{equation} where $\delta_{z}$ = ($z_{b}$-$z_{s}$) is $z_{b}$ and $z_{s}$ the photometric and spectroscopic redshift estimate, respectively. Apart from the scatter, we also computed the mean ($\mu$) value for the error distribution to explore unwilling systematic biases at particular redshifts. The fraction of catastrophic errors (defined in Equation $\ref{outlieq}$) was also computed to understand the expected contamination of our photo-z samples when looking for cluster galaxies (see Section $\ref{ClusterMembers}$). In this work we defined as catastrophic outliers those galaxies with an error in their redshift estimation larger than five times the typical error distribution ($\sigma_{NMAD}$):  

\begin{equation}
\,\,\,\,\,\,\,\,\,\,\,\,\,\,\,\,\,\,\,\,\,\,\,\,\,\,\,\,\,\,\,\,\,\,\,\,\,\,\,\,\,\,\,\,\,\,\,  \eta = {\left | \delta_{z} \right | \over 1+z_{s}} > 5 \times \sigma_{NMAD} 
\label{outlieq}
\end{equation}We devote the following subsection to the discussion of the different methods used to characterize the quality of our photo-z estimations, using real and simulated data.

\subsection{Photometric redshift quality.}
\label{photozquality}
Galaxies at low redshift ($z$<1.0) preserve the majority of their most distinct spectral features within the optical window. In such circumstances, main uncertainties in the photo-z estimations in optical surveys come from the available wavelength resolution (number and type of passbands) and from the limited photometric depth of the observations. These two factors act in the same direction and may cause the mapping of the SED and of the redshift space uncertain. Keeping that idea in mind, in this section we present two parallel analyses. First, in Section \ref{accspeczdata} we quantify the photo-z performance reached by both the J-PLUS and the SDSS (DR12) surveys for the spectroscopic sample of $\sim$300 cluster galaxies presented in Section \ref{data}. Taking into account that both surveys have a similar photometric-depth, this comparison proves the benefit of increasing the wavelength resolution from 5 (SDSS) to 12 (J-PLUS) filters. Second, taking into account that the spectroscopic sample is mainly composed by bright galaxies ($r_{SDSS}$<18), we develop a set of simulations to extend the sample to fainter magnitudes ($r_{SDSS}$<20) and test the expected performance of both surveys for galaxies at low S/N levels. This analysis is presented and discussed in Section \ref{accsimuldata}.

\vspace{0.2cm}

\subsubsection{Performance on real data.}
\label{accspeczdata}
\begin{figure}
\centering
\includegraphics[width=9.cm]{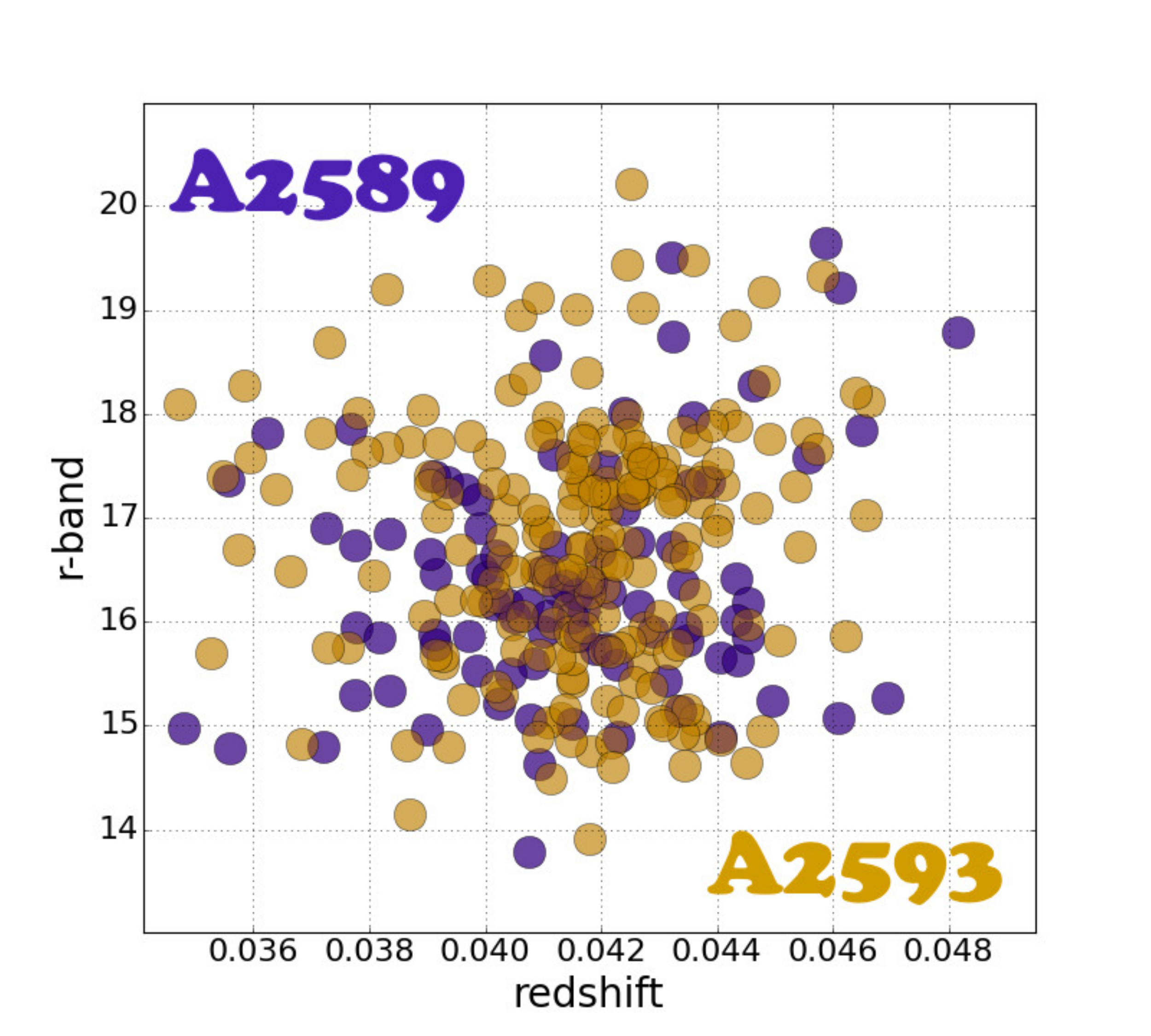}
\caption[spectroscopic sample.]{Spectroscopic redshift sample. The figure shows the sample composed by 297 confirmed galaxy cluster members used in this work to characterize the quality of the J-PLUS photo-z.}
\label{speczsample2}
\end{figure}

\begin{figure*}[tbp]
\centering
\includegraphics[width=9.4cm]{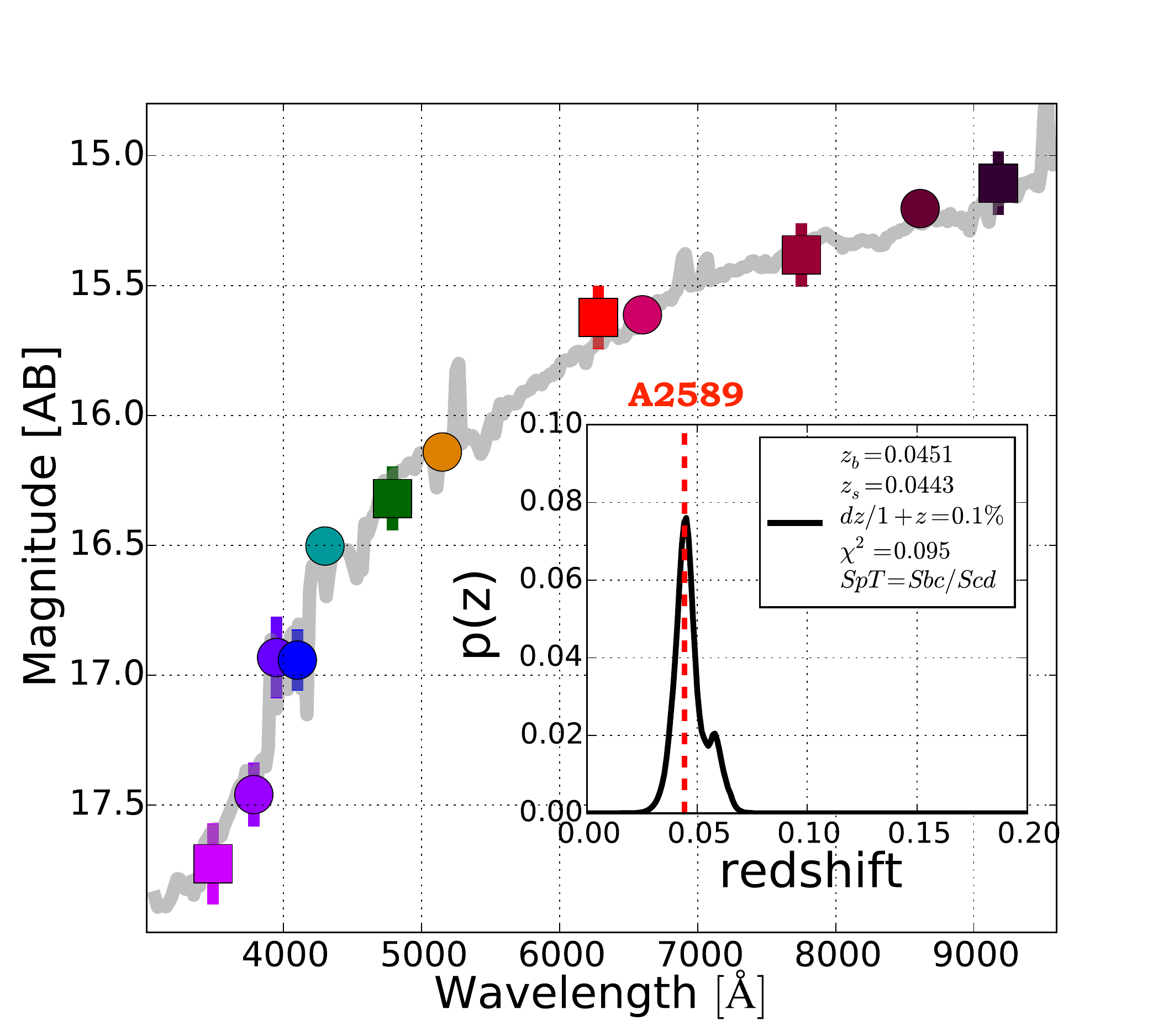}
\includegraphics[width=8.65cm]{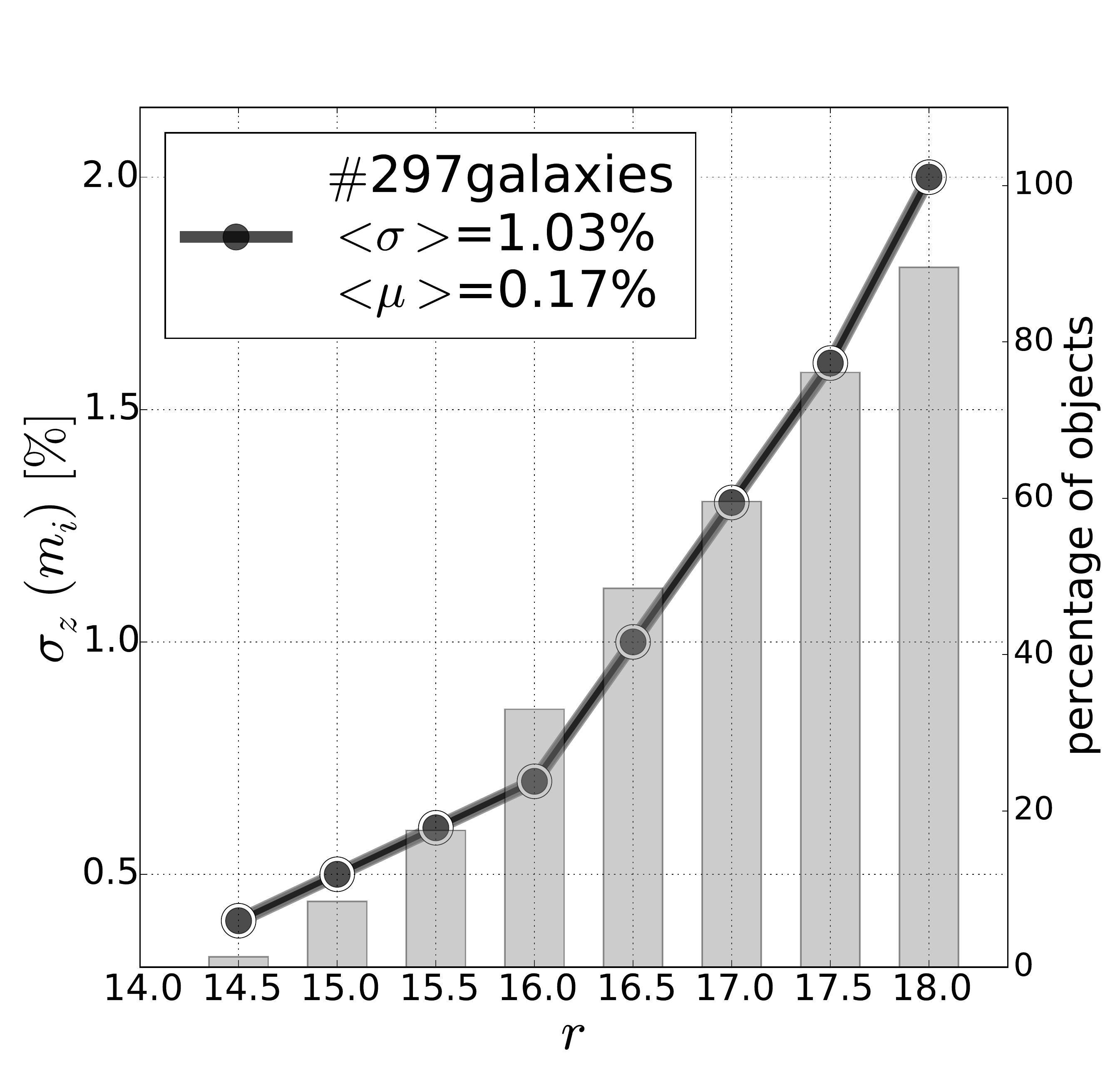}
\caption[photoz]{Photo-z performance of the J-PLUS filter system. Left: The figure shows an example of the SED-fitting for an early-type cluster galaxy. Inner-panel shows the PDF computed by the \texttt{BPZ} code. Right: the figure shows the obtained accuracy as a function of the cumulative $r$ magnitude. Gray vertical bars represent the fraction of the galaxies per magnitude bin. Globally, this sample yields an accuracy of $\delta_{z}$/(1+$z$) $\sim$1.0\% with an averaged magnitude $<r>$=16.6. A precision of $\delta_{z}$/(1+$z$) = 0.005 is obtained for the 177 galaxies brighter than magnitude $r<$17, showing the enormous potential of this technique to study the nearby galaxies.}
\label{photoz1}
\end{figure*}

Initially, we used the spectroscopic sample composed by 297 galaxies confirmed as cluster members (see Figure $\ref{speczsample2}$) to characterize the expected performance of our J-PLUS photo-z estimations. These results, summarized in Table $\ref{dzvsmagtable}$ and the right panel of Figure $\ref{photoz1}$, show the potential of this new generation of multiband photometric surveys to derive accurate photo-z estimations. Globally, this sample yields an accuracy of $\delta_{z}$/(1+$z$) $\sim$1.0\% with an averaged magnitude $<r>$=16.6. An accuracy of $\delta_{z}$/(1+$z$) $\sim$0.52\% is yield for the 177 galaxies brighter than magnitude $r<$17. The fraction of catastrophic outliers ($\eta$) is always below 2\% and the bias ($\mu$) is always smaller than 0.2\%. An example of the SED-fitting for a cluster member is also illustrated in the left panel of Figure $\ref{photoz1}$. As mentioned in Section \ref{observations}, photo-z estimates have already been done for the A2589 cluster using a similar 15 intermediate-band system from the Beijing-Arizona-Taiwan-Connecticut (BAT) survey. As stated in SF11, they reach a precision of $\delta_{z}$/(1+$z$)=0.0077 for the galaxies brighter than $r<$17 in that particular cluster. These results are certainly similar to those obtained with our filter system using a similar number of passbands, reassuring the reliability of our estimates. 

\vspace{0.2cm}

Then we estimated the photo-z precision on the same sample reached by the SDSS/DR12 data using only five broadband filters. To do so, we used the magnitudes from SDSS/DR12 presented in Section \ref{sdss2jplus} and run \texttt{BPZ} using the same configuration as that used for J-PLUS. Globally, SDSS photometry yields an accuracy of $\delta_{z}$/(1+$z$) $\sim$2.5\% for the whole sample and an accuracy of $\delta_{z}$/(1+$z$) $\sim$1.9\% at magnitude rSDSS$<$17. The fraction of catastrophic outliers is typically 1\% and the bias is always $<$1\%. These results indicate that the J-PLUS photo-z may surpass those from SDSS by a factor of 2 using a standard 5 broadband filter system, in the nearby Universe.   

\begin{table}
\caption{{\small Photo-z quality. The table summarizes the obtained photo-z accuracy as a function of the $r$ magnitude, for both the J-PLUS ($J$) and the SDSS ($S$) surveys, on a sample of $\sim$300 spectroscopically confirmed cluster galaxies. The different parameters are defined in Section \ref{photozs}.}}
\begin{center}
\label{dzvsmagtable}
\begin{tabular}{|c|c|c|c|c|c|c|c|c|}
\hline
Mag  & $\sigma_{z}^{J}$ & $\mu_{z}^{J}$ & $\eta ^{J}$ & $\sigma_{z}^{S}$ & $\mu_{z}^{S}$ & $\eta^{S}$ & \#  \\ 
\hline
  [AB] & (\%) & (\%) & (\%) & (\%) & (\%) & (\%) & (\%) \\
\hline
$r$ $<$ 15 & 0.3 & 0.0 & 0.00 & 1.7 & 0.7 & 2.9 & 8  \\
$r$ $<$ 16 & 0.4 & 0.1 & 0.00 & 1.9 & 0.6 & 0.9 & 32 \\
$r$ $<$ 17 & 0.5 & 0.0 & 0.00 & 1.9 & 0.6 & 0.5 & 59 \\
$r$ $<$ 18 & 0.9 & 0.1 & 0.00 & 2.4 & 0.5 & 0.3 & 88 \\ 
$r$ $<$ 19 & 1.0 & 0.2 & 0.35 & 2.5 & 0.5 & 0.7 & 96 \\ 
$r$ $<$ 20 & 1.1 & 0.2 & 1.69 & 2.5 & 0.5 & 0.9 & 100 \\ 
\hline
\end{tabular}
\end{center}
\end{table} 

\subsubsection{Performance on simulated data.}
\label{accsimuldata}

\begin{figure}[tbp]
\centering
\includegraphics[width=8.5cm]{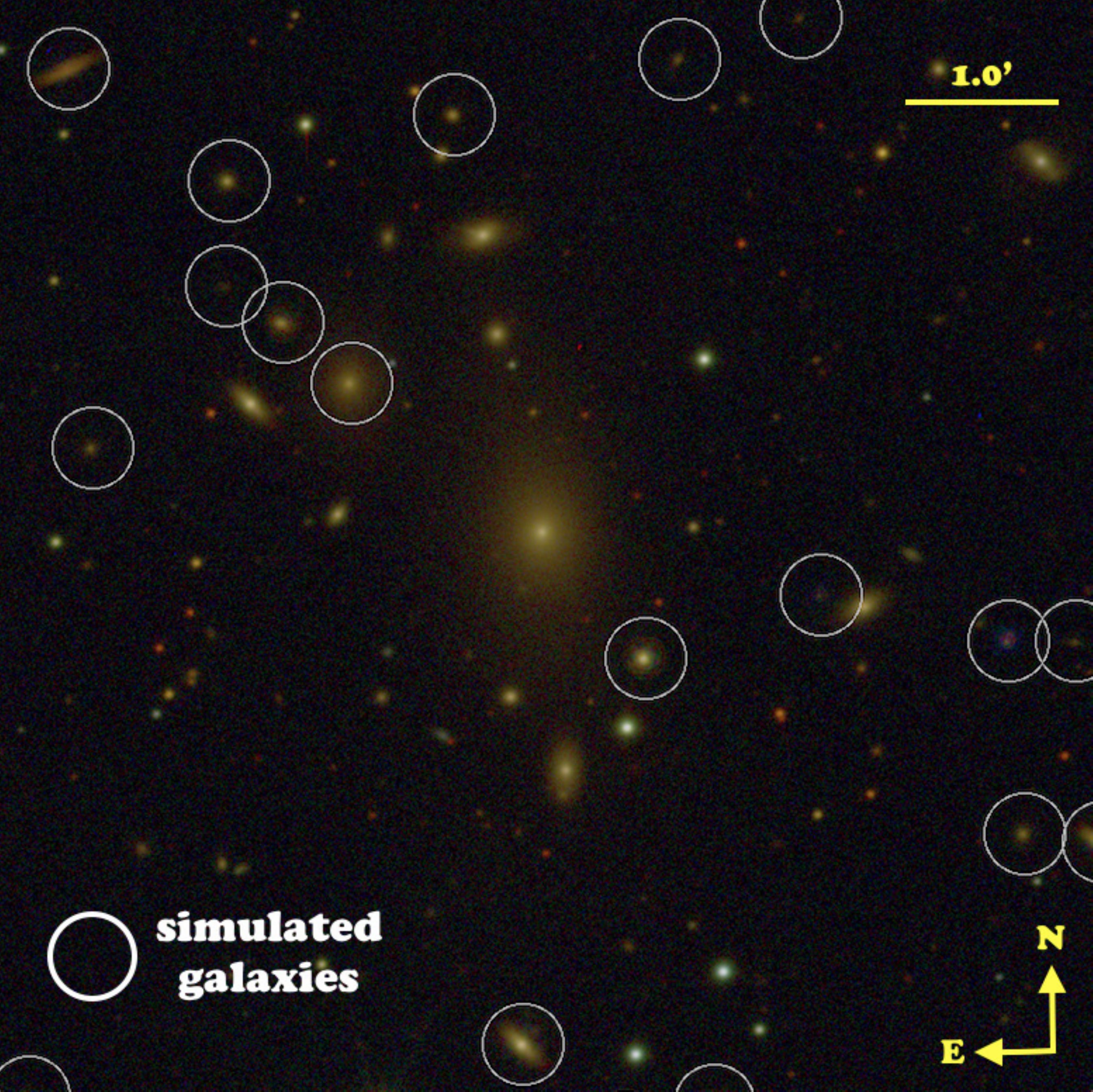}
\includegraphics[width=9.cm]{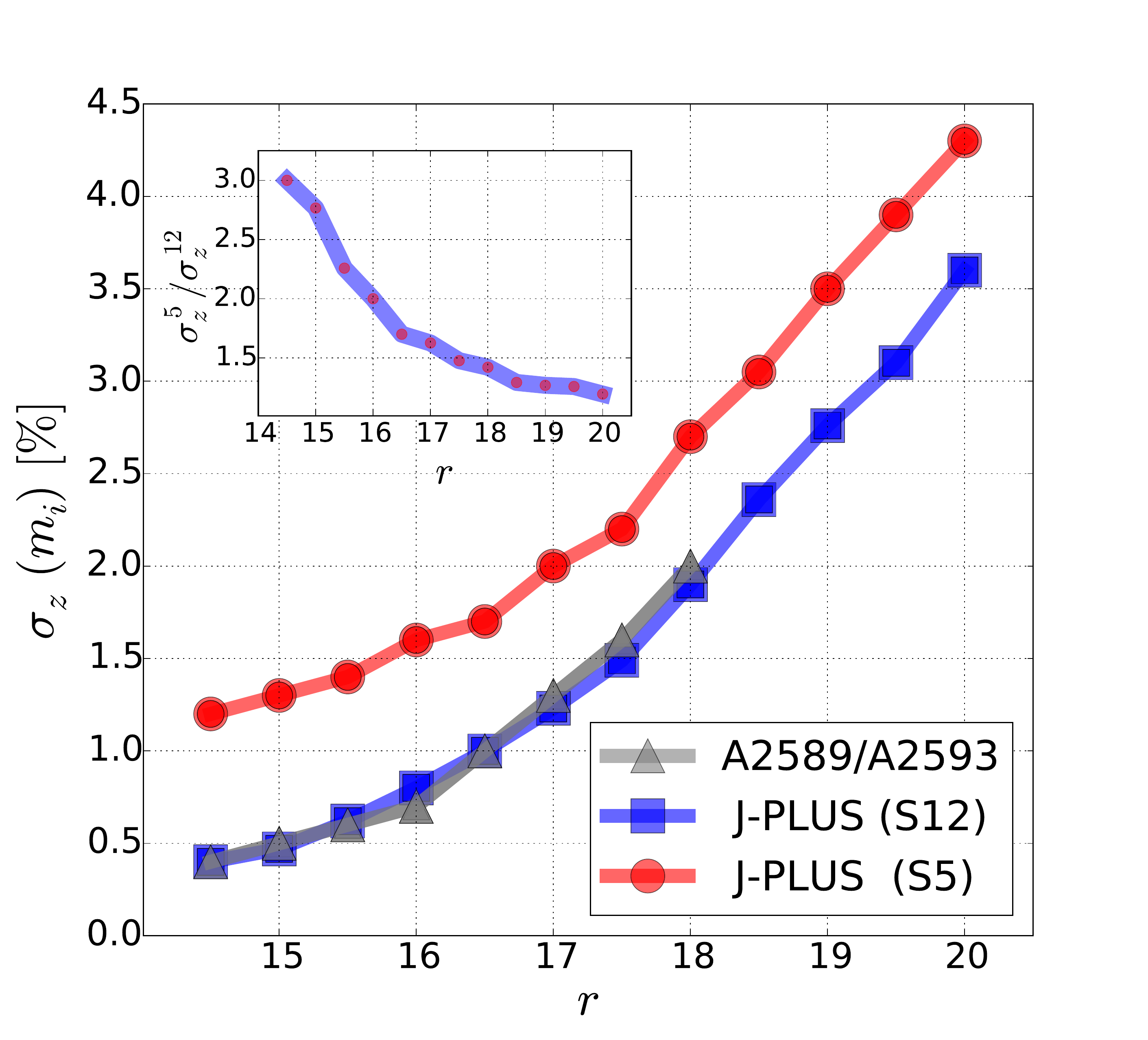}
\caption[Simulated cluster members]{Cluster member simulations. Upper panel shows an example of the simulated images where galaxy models are injected at different positions and magnitude ranges. Lower panel shows an example of the photo-z results obtained from the $\sim$10k simulated galaxies using both 12 and 5 filters, spanning a magnitude range from 14 < $r_{SDSS}$ < 20. The results indicate that there is a net improvement when the wavelength resolution is increased at a similar depth.}
\label{speczsimul1}
\end{figure}

As mentioned before, due to the fact that the spectroscopic sample was mainly dominated by bright and high S/N galaxies, we wanted to explore the representativeness of this particular dataset with respect to the expected performance for the J-PLUS survey on nearby galaxies in clusters. To do this exercise, we decided to design a set of simulations to extend our control sample to a much lower S/N. Rather than generating mock catalogs with expected colors for galaxies observed through our filter system, we preferred to create mock images where galaxies with known properties were injected into the original images. This approach has the benefit of preserving all the photometric complexity encoded in real data (such as noise, PSF variations and/or pixel-correlations). The top panel of Figure $\ref{speczsimul1}$ illustrates an example of one such simulated image.

\vspace{0.2cm}

We rely on the \texttt{CHEFs} software\footnote{The software utilizes a library of Chebyshev-Fourier mathematical functions in a nonparametric fashion to efficiently model the light surface distribution of galaxies irrespective of their morphologies.} (\citealt{2012ApJ...745..150J}; \citealt{2015MNRAS.453.1136J}) to model all the spectroscopic galaxies in the 12 bands. These models were stored preserving a reference magnitude, redshift and cluster name information. In order to expand the magnitude range from 14$<$r$<$22, we artificially flux-scaled the models to fit the new reference magnitudes. These new flux-scaled galaxies were randomly injected across the images in positions where no galaxies were reported by the \texttt{SExtractor} segmentation-maps to avoid overlapping effects that have nothing to do with the impact of noise on photo-z estimations. In order to increase as much as possible the statistics of this analysis without changing the noise properties of our images, we restricted the amount of models to inject in every iteration to 1K, repeating the exercise several times until a large and meaningful statistics was compiled\footnote{The T80Cam imager has 9.2kx9.2k pixel array. Therefore even if these 1K galaxies were 500-pixel-size each (10 times larger than the PSF-size), they would still represent less than 1\% of the total image area.}.

\vspace{0.2cm}

Over this new set of images, we ran the same pipeline presented in Section $\ref{photometry}$, computing a new multiband photometry and photo-z estimations for those galaxies. The whole process was repeated several times in order to compile a final catalog of $\sim$10k galaxies. As expected, the photo-z accuracy had this time a larger variation than that obtained using the spectroscopic sample, spanning a range of values from $\delta_{z}$/(1+$z$) $\sim$0.5\% to $\delta_{z}$/(1+$z$) $\sim$3.0\% at different magnitude bins. Although there may seem to be at first glance, there is no contradiction between the former results obtained with the (real) spectroscopic sample and the ones with the simulated sample. While galaxies in the spec-z sample are mainly concentrated at brighter magnitudes $<$r$>$=16.6 (see Figure $\ref{speczsample2}$), the simulated sample homogeneously spread the magnitude range from 14$<$r$<$22. As seen in the bottom panel of Figure $\ref{speczsimul1}$, simulated galaxies with magnitudes r$\sim$16 also reach a value of $\delta_{z}$/(1+$z$) $\sim$1.0\%. The differences between the two samples rely solely on the different magnitude distributions of the galaxies. This fact stresses the importance of using large samples of spec-z galaxies. 

\vspace{0.2cm}

Finally, we take advantage of these simulations to compare again the expected accuracy yielded by a SDSS-like survey when decreasing  the number of bands from 12 to 5. To do so, we re-ran $\texttt{BPZ}$ using only the 5 broad-bands in our filter system ($u$, $g$, $r$, $i$, \& $z$). As seen in Figure $\ref{speczsimul1}$, the expected degradation in the photo-z performance is observed when using this lower wavelength-resolution. We note that even at high S/N level, the SDSS-like surveys cannot surpass a certain precision in the photo-z estimates. This is due to the fact that the limited wavelength resolution provided by the standard five broadbands, causes a degeneracy in the color-redshift space. These uncertainties dominate over the photometric signal from images. The results illustrated in Figure $\ref{speczsimul1}$ show again how the J-PLUS(12) photo-z estimations can surpass those from J-PLUS(5). Specially at the bright side where the narrowband filters still preserve a large S/N.

\vspace{0.2cm}

Two specific points of interest can be extracted from these simulations: firstly, the average photo-z accuracy provided by the  spectroscopic sample (presented in Section \ref{accspeczdata}) is realistic and representative of the expected performance of the J-PLUS photo-z. However, it is important to keep in mind that larger, more heterogeneous and fainter samples (as those proposed in our simulations) are needed to asses realistic expectations for the J-PLUS photo-z in nearby galaxy clusters at faint magnitudes. Secondly, there is a significant improvement in the photo-z performance whit the increasing of the wavelength resolution. Up to a factor of 2.5 improvement between the J-PLUS(5) and the J-PLUS(12) photo-z can be achieved at high S/N levels. 

\subsection{Photometric zero-point refinements.}
\label{photozpcal}
When using photo-z codes based on template fitting, if the redshift of the galaxies are known, it turns out possible to compare the expected colors (fluxes) from a library of galaxy models with those observed from the real galaxies. If the photometry is accurate and the library of models reliable (in terms of both calibration and completeness), the dispersion among expected and observed colors is supposed to be caused by the noise from images. If this is true, the ratio among fluxes can be approximated to a normal distribution with mean equal 1 ($\mu=1$) and a dispersion proportional to the background noise. Statistical deviations from that unicity are, therefore, assumed to be instrumental zero-point offsets that can be included in the original photometry. The procedure is iteratively repeated until convergence is reached (\citealt{2006AJ....132..926C}; \citealt{2014MNRAS.441.2891M}).  

\vspace{0.2cm}

In this work we used the spectroscopic sample (see Section $\ref{observations}$ and Figures \ref{speczsample1} and \ref{speczsample2}) to refine our initial zero-point estimates. To preserve the absolute calibration of our photometry, we forced \texttt{BPZ} to not change the fluxes from the ($g$, $r$, $i,$ \& $z$) broadbands. This guarantees that our final photometry will remain tightened to that of SDSS. The so-derived corrections are applied to the photometry before computing photo-z estimates for the whole catalog. The corrections were always small ($<$5\%) compatible with the noise level and the color differences from the definition of photometric apertures. We show in Figure \ref{photoz1} an example of the final J-PLUS photometry for an early-type cluster galaxy. As discussed in Section \ref{photometry}, this includes aperture-matched PSF-corrected magnitudes across bands, an empirical noise re-estimation and a zero-point refinement based on photo-z (this Section).

\section{Statistical cluster member identification.}
\label{ClusterMembers}

One of the main motivations of this work is to study the feasibility of using the J-PLUS data to revisit membership analysis in nearby clusters of galaxies. As a test-case, we used accurate multiband photometry and Bayesian photometric redshifts (Section \ref{photozs}) to look for potential new cluster members in (and in between) the galaxy clusters A2589 \& A2593. In the following sections, we describe the methodology we have adopted to flag potential candidates using PDFs (Section \ref{IPDZ}), the different analysis we have done to quantify the reliability of this technique (Section \ref{IPDZtesting}) and the results we obtained after applying this methodology to the general catalog presented in Section \ref{spatialdistribution}.

\vspace{0.2cm}

As mentioned in Section \ref{observations}, the identification of new members in the galaxy cluster A2589 has also been addressed in SF11. Although similar in scope, in this work we have adopted a different approach to address this problem. As explained in SF11, photo-z estimates are treated as point-estimates rather than as complete distribution functions in a bi-dimensional (color - redshift) space. It is worth noting that galaxies with a high S/N and good spectral coverage may have a photo-z PDF close to a Gaussian function, with a dispersion similar to the photo-z error distribution. However, these PDFs usually depart from Gaussianity at low S/N levels, showing complicated multimodal shapes (i.e., several peaks). Under such circumstances, the treatment of photo-z as simple point-estimates may lead to potential biases in the study of clusters of galaxies. An example of such scenario would be a galaxy with a significant fraction of its probability in redshift within the cluster interval but with the maximum of its distribution not laying within such interval. Considering that we are precisely interested in conducting membership analysis for both bright and faint galaxies, we have preferred to use all the information stored in these PDFs without making the strong assumption of collapsing the entire redshift distribution of every galaxy to a single value. 

In addition, in SF11 a fixed and rigid redshift interval is adopted for the computation of membership, using a standard 3$\sigma$-clipping with respect to the cluster redshift ($z_{ph}<$ abs($z_{cl}\pm 3\times\sigma_{z}$)), where $\sigma_{z}$ would be the observed photo-z error from a spectroscopic sample of bright galaxies. Although this definition may be accurate for galaxies with magnitudes similar to those from the spectroscopic sample, there is no guarantee that such uncertainty ($\sigma_{z}$) may be representative for galaxies at fainter magnitudes (r$>$18.0). As discussed in the next section, we have adopted a different approach in this work, relying on simulations to describe the real transformation of PDFs as a function of the apparent magnitude. Eventually, these empirical distributions are used to define magnitude-dependent redshift intervals over which to integrate the membership probabilities of individual galaxies in our fields. 

\subsection{Integrated probability distribution function in redshift.}
\label{IPDZ}
For every galaxy in our catalog, the $\texttt{BPZ}$ code provides a complete PDF in a bi-dimensional (redshift versus spectral-type) space. Based on these distributions, it turns out possible to identify potential new cluster members as those galaxies with a significant fraction of their probabilities within the cluster interval. In order to be able to properly define this interval over which to integrate individual probabilities, we have decided to rely on simulations. As explained in Section $\ref{accsimuldata}$, a number of spectroscopically confirmed cluster galaxies were first modeled, flux-scaled and reinjected in our images, recomputing afterward their multiband photometry and photometric redshifts with the same pipeline presented in Section \ref{colorpro}. Due to the fact that these simulations are solely based on cluster galaxies, it renders possible to understand how the PDFs of such galaxies change as a function of the magnitude (i.e., S/N). Likewise, as these simulations include a large number of models, it turns out possible to accurately derive representative PDFs for cluster galaxies at different magnitude-bins. 

Following this premise, we generate ``master'' PDFs at different magnitude-bins as follows. From our catalog, we initially select all sources classified as galaxies within a certain magnitude-bin $m_{j}$. Later on, we marginalize and normalize every $i$th-PDF over the spectral-types using the following equation: 

\begin{equation}
P_{i}(z)  = {\int_{z_{1}}^{z_{2}}}{\int_{T_{1}}^{T_{2}}} p_{i}(z,T) \,dT \,dz = {\int_{z_{1}}^{z_{2}}} p_{i}(z) \,dz = 1  
\label{normedPDZ}
\end{equation} where $P_{i}(z)$ represents the collapsed probability distribution function in redshift for the $i$th-galaxy, $z_{1}$ \& $z_{2}$ the minimum and maximum redshift values allowed to compute our photo-z estimates and $T_{1}$ \& $T_{2}$ the first and last template in the $\texttt{BPZ}$ library of galaxy models. These probabilities are finally combined and normalized to generate the magnitude-dependent Master PDFs. It is worth noting that since these Master distributions correspond to the typical spread of individual PDFs for cluster galaxies at a given magnitude-bin, they represent the natural redshift interval over which to integrate individual probabilities. Finally, for the sake of clarification, we rename these distributions as the cluster intervals (hereafter $CI$), after imposing the usual normalization factor:

\begin{equation}
CI_{j} (z)  = {\int_{z_{1}}^{z_{2}}} [ \sum_{i} P_{i}(z) ] \, dz = 1
\label{normedPDZ2}
\end{equation} where $CI_{j} (z) $ would correspond to the cluster interval in redshift space for the $j$th magnitude-bin. 

\vspace{0.1cm}

We compared the resemblance of the Master PDFs derived from simulations with those obtained directly from real data using a Kolmogorov-Smirnov test. Due to the limited number of available galaxies in the spectroscopic sample at the faintest magnitude-bins, we compare only the obtained distributions for galaxies brighter than magnitude $r=$18 ($\sim$90\%; see right panel of Figure $\ref{photoz1}$). This exercise shows that both distributions are $<$5\% different, assuring the validity of using simulated data to describe the PDFs of real galaxies. 

\vspace{0.1cm}

Finally, we defined the integrated probability in redshift (hereafter $IPDz$) for each galaxy using Equation $\ref{ipdz}$. This quantity corresponds to the fraction of the total probability of the $i$th-galaxy (with a magnitude $m_{j}$) within the previously defined cluster interval $CI_{j}(z)$:

\begin{equation}
IPDz_{i} = P(z_{i} \cap z_{cl})  = {\int_{z_{1}}^{z_{2}}} {p_{i}(z) \, CI_{j}(z) \,dz \over p_{i}(z) \,dz}
\label{ipdz}
.\end{equation}

Based on this definition for the $IPDz$, we selected as potential cluster member candidates those galaxies with a $IPDZ$ $\geq$0.5; that is, those with a probability within the cluster interval larger than that on the field (foreground + background). In addition, we imposed a minimum S/N $\ge$ 10, according to the detection image to remove galaxies with very poor photo-z estimates (noisy $PDF$). After imposing all these conditions, we find a total of $\sim$440 potential cluster members in the entire field; $\sim$210 members in pointing $P01$ (A2593), $\sim$60 members in pointing $P02$ and $\sim$140 members in pointing $P03$ (A2589). After excluding all cluster galaxies already included in the spectroscopic sample, we find as much as 70 new potential members in pointing $P01$ and $\sim$60 in pointing $P03$. Our results are in a good agreement with those from $SF11$ for the A2589 cluster ($P03$). In SF11 as much as 174 galaxies with magnitudes brighter than V=20 are classified as cluster members; being 110 new candidates to be cluster members. 

\subsection{Reliability of this technique.}
\label{IPDZtesting}

In order to quantify the reliability of this technique finding potential new cluster members, we made use of both the spectroscopic sample of cluster galaxies presented in Section \ref{observations} and the simulated data presented in Section \ref{accsimuldata}. We defined the completeness ($C$) factor as the ratio between the number of galaxies classified as cluster members ($IPDZ\geq$0.5) over the total number of galaxies in a particular magnitude bin. Based on the spectroscopic sample, we retrieved a 97$\%$ completeness for the whole sample. Interestingly, we find that $\sim$3\% of the bright galaxies (spectroscopically) confirmed as cluster members got a probability $IPDZ<<$0.5. After inspecting their $PDF$, we found that they had narrow and single-peak distributions (suggesting a secure photo-z) but incompatible with the cluster redshift. This contradiction could be explained either due to problems when cross-matching (photometric and spectroscopic) catalogs or due to misclassified cluster members in the spectroscopic sample.  Likewise, we used the simulated sample to characterize the expected $C$ of our sample. For galaxies with magnitudes brighter than $r<$17, we expect a completeness factor of $C$=1.0 (i.e., 100\%). For galaxies with magnitudes $17<r<19$ a $C\sim$0.8 and for galaxies with magnitudes $19<r<20$ a $C\sim$0.7. 

\vspace{0.2cm}

In the same manner, we estimated the expected fraction of misclassified ($M$) cluster members in our fields. Due to the fact that the spectroscopic sample used above to compute $C$ does not contain field galaxies, and that it is not possible to define a control field to calibrate our integrated probabilities at the time of this paper, we decided to adopt a different strategy. We made use of the Early Data Release (EDR)\footnote{The EDR is presented in \citealt{Cenarro18}.}available at the time of this paper ($\sim$200 deg$^{2}$), to compile a sample of $\sim$600 galaxies with spectroscopic redshifts measured by SDSS/DR9 (internal communication). We excluded all galaxies with redshifts within the cluster range (see Figure \ref{speczsample2}) to create a ``field'' sample, with an average magnitude of $<r>\sim$17.0 and a redshift range 0.001$<z<$1.0. We ran $\texttt{BPZ}$ with the same configuration used in this work and derived new $PDF$. Finally, we computed the fraction of the spectroscopic galaxies classified as potential cluster members according to our methodology. For galaxies with magnitudes brighter than $r<$16, we find a $M$ $\sim$0.4\% (2/520); for galaxies with a magnitude in between $16<r<18$ a $M$ $\sim$4\% (44/1100); for galaxies with a magnitude in between $18<r<20$ a $M$ $\sim$9\% (92/1100). 

We emphasize that even if the spectroscopic sample used to compute $M$ is similar in terms of the magnitude to that used to compute $C$, the redshift distribution was rather peculiar. Therefore, the so estimated values of $M$ should be handle with care and not freely extrapolated to the general survey. A further analysis on the issue will be done in the future when other cluster fields (with abundant and heterogeneous spectroscopic information) become available in the survey.

\subsection{Spatial distribution of potential new cluster members.}
\label{spatialdistribution}

\begin{figure*}[tbp]
\centering
\includegraphics[width=18.5cm]{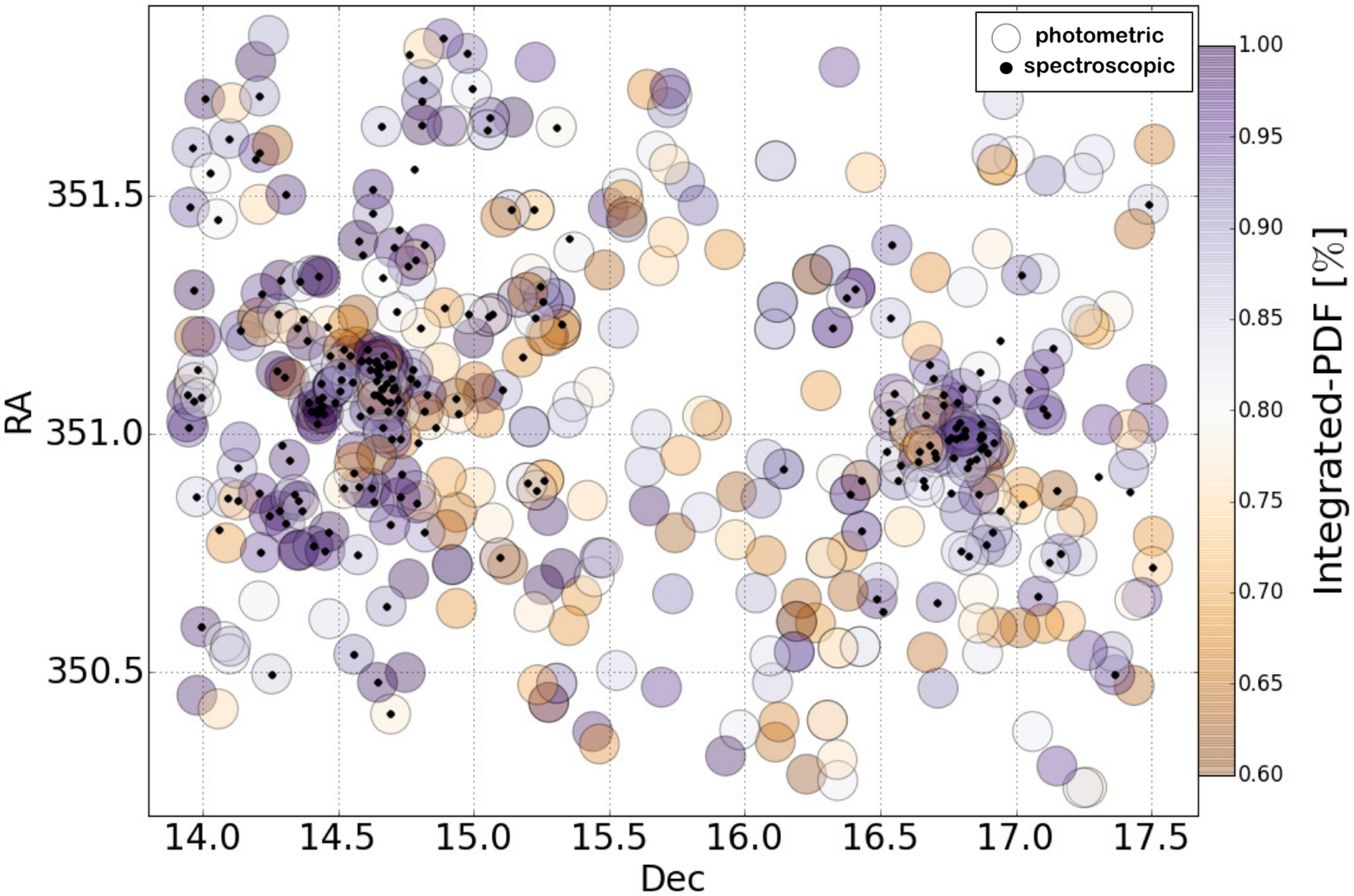}
\caption[Spatial distribution of potential new cluster members.]{All the identified galaxies as potential cluster members based on the integrated probability distribution function ($IPDF$) technique. As explained in Section \ref{IPDZ}, these probabilities are integrated within an interval empirical defined for the cluster galaxies based on simulations. As much as 170 galaxies are flagged as potential new cluster members across the entire field. The spatial distribution of the potential new cluster members may suggest that A2589 \& A2593 could be two connected systems although there is not any clear filamentary structure linking both clusters.} 
\label{ipdzplot}
\end{figure*}

After estimating the cluster membership for all galaxies in our fields, we have proceeded to study the spatial distribution of all potential cluster member candidates in a RA-Dec-z space. As shown in Figure \ref{ipdzplot}, where the integrated probability is color-coded\footnote{A value of $IPDZ$>0.6 was arbitrarily chosen to increase the color contrast and facilitate the visualization of this figure.}, the majority of these candidates were spread over the clusters at very similar positions to those occupied by the spectroscopically confirmed members. Interestingly, we also observed a clear excess of galaxies with high $IPDZ$ (>80\%) in the intermediate pointing ($P02$) resembling a connecting structure between the clusters. 

\vspace{0.2cm}

Finally, in order to explore the impact of the photo-z resolution on the identification of new cluster members, we decided to repeat the estimation of the integrated probabilities ($IPDZ$, presented in Section \ref{IPDZ}) but using solely the five broadband filters in common with the SDSS survey; that is, removing the seven narrow-bands from our filter system. To do so, we reran the $\texttt{BPZ}$ code using exactly the same configuration as before, computing a new $PDF$ for each galaxy. Interestingly, this test showed that a large fraction of galaxies previously classified as potential new members were now classified as field galaxies. Likewise, the observed over-density in between the two systems using the 12 bands almost vanished when using only five. The result emphasizes the importance of improving the resolution of photo-z estimations and its capability to rise overlooked faint structures in the nearby Universe. 

\section{Discussion.}
\label{discussion}  

As discussed in Section \ref{spatialdistribution}, the spatial distribution of the potential new cluster members may suggest that the clusters A2589 \& A2593 could be two connected (rather than isolated) systems. In order to test this hypothesis, we initially inspected the spatial distribution of the X-ray emission from the Rosat All-Sky Survey (RASS; \citealt{1999A&A...349..389V}). Although both clusters were easily identifiable in the X-ray maps, we could not find any clear sign of a filament linking both clusters. This result suggests that a hypothetical filament may be quite poor on diffuse warm gas. Later on, we performed a study of the spatial distribution of all potential members, similar to that presented in \cite{2017MNRAS.466.2614M}, looking for any substructures in the field connecting the clusters. Again, this analysis suggested that a simple two-component (two clusters) rather than a three-component (two clusters + a filament) system was clearly favored. Based on this evidence, we conclude that there is not a clear evidence of any connecting structure linking the systems. A possible explanation may be that the observed excess of candidates in between the clusters is due to a overlap in the spatial distributions of galaxies in these systems.

\vspace{0.2cm}

Although the physical distance between both clusters ($\sim$7 Mpc) excludes a merging scenario, we cannot rule out that the structure is collapsing, falling toward each other and, therefore, not showing any clear evidence of this interaction yet in the hot X-ray emitting gas at this stage. Despite the majority of the spectroscopically confirmed cluster members are successfully identified by the method proposed in this work, the restricted available area for this analysis (Phase-3 Verification data) makes impossible at this point to confirm if the observed overdensity between A2589 \& A2593 clusters is real (an overlap of both clusters) or just an artificial projection of a large structure of background and foreground galaxies falling into the clusters. To tackle this question, it may be necessary to extend our observations to a larger area, so we can confirm whether or not the overdensity is particular of the line that unites both clusters and to quantify the net contamination from fore- or background galaxies entering the cluster redshift range due to the limited redshift resolution of our photo-z estimates. These data will be available once the J-PLUS survey has accomplished its observations.

\section{Summary}
\label{summary}   
In this work we have used multiband imaging from the Phase-3 verification data of the J-PLUS survey (\citealt{Cenarro18}) to look for potential new cluster members in the two nearby galaxy clusters A2589 ($z$=0.0414) \& A2593 ($z$=0.0440), using accurate Bayesian photo-z estimations derived from a 7 narrow + 5 broadband filter system. The optimize pipeline for clusters of galaxies adopted in this work includes a PSF-corrected photometry with a specific definition of the apertures that enhances the S/N in the bluest filters, an empirical re-calibration of the photometric uncertainties and accurate photometric upper-limit determinations. A comparison of our photometry with that from the SDSS/DR12 shows a good agreement with typical dispersion of $\sim$10\% at a magnitude $r$=18 and negligible bias. In order to decontaminate our catalogs from stars, we apply a star/galaxy classification method based on random forest, combining multiband and morphological information (Costa-Duarte et al., in prep).

The pipeline presented in this work provides $\delta_{z}$/(1+$z$) = 0.01 accurate photo-z when compared to a sample of $\sim$300 spectroscopically confirmed cluster galaxies, with an average magnitude of $<r>$=16.6. A precision of $\delta_{z}$/(1+$z$) = 0.005 is obtained for the 177 galaxies brighter than magnitude $r<17$, showing the enormous potential of the J-PLUS data for galaxies with a high S/N level. In order to test the expected performance of our photo-z estimations for faint galaxies, we designed a set of simulations in which real cluster galaxies are modeled, flux-scaled and re-injected in our images. We rely on these mock images to estimate the expected uncertainties in the photo-z estimations at a low S/Ns and to compare the benefit of extending classical five broad-band surveys (SDSS-like) to twelve bands (J-PLUS). We find a net improvement at all magnitudes, reaching up to a factor of two improvement when the S/N in the narrow-band filters is high enough. Taking advantage of these simulations, we derived empirical $PDF$ as a function of the magnitude. These distributions represent the real uncertainties in the redshift-space for our photo-z estimations. 

Likewise, we used these master $PDF$ to look for potential new cluster members in and in between the A2589 \& A2593 clusters. We find as many as 170 potential new members across the entire field, in good agreement with the results obtained in SF11 for the A2589 cluster. The spatial distribution of these candidates may suggest a connection between the systems, where surrounding galaxies may be entering the main clusters. However, neither the distribution of X-ray emission nor the spatial distribution of the potential new members between these systems can confirm this scenario although they do not exclude it neither. One other scenario may be that there is a gentle ongoing interaction among the two systems, forming perhaps an eventual supercluster structure. Due to the restricted available area for this analysis (Phase-3 Verification data), we cannot confirm at this point if the observed over-density between the A2589 \& A2593 cluster is real (due to an overlap in the spatial distribution of galaxies in these systems) or just an artificial projection due to the moderate blurring of our photo-z estimates tracing the real underneath distribution of galaxies. Nevertheless, these results show the potential of new multiband photometric redshift surveys to revisit theories of cluster formation and evolution in the nearby Universe.     

In addition to the present paper, the J-PLUS EDR and science verification data were used to analyze the globular cluster M15 (\citealt{Bonatto18}), study the H$\alpha$-emission (\citealt{Logrono18}) and the stellar populations (\citealt{SanRoman18}) of several local galaxies, and compute the stellar and galaxy number counts up to $r = 21$ (\citealt{LopezSanjuan18}).

\begin{acknowledgements}
We acknowledge the OAJ Data Processing and Archiving Unit (UPAD) for reducing and calibrating the OAJ data used in this work. Based on observations made with the JAST/T80 telescope at the Observatorio Astrof\'isico de Javalambre, in Teruel, owned, managed, and operated by the Centro de Estudios de F\'isica del Cosmos de Arag\'on (CEFCA). Funding for the J-PLUS Project has been provided by the Governments of Spain and Arag\'on through the Fondo de Inversiones de Teruel, the Arag\'on Government through the Reseach Groups E96 and E103, the Spanish Ministry of Economy and Competitiveness (MINECO; under grants AYA2015-66211-C2-1-P, AYA2015-66211-C2-2, AYA2012-30789 and ICTS-2009-14), and European FEDER funding (FCDD10-4E-867, FCDD13-4E-2685). AM acknowledges the financial support of the Brazilian funding agency FAPESP (Post-doc fellowship - process number 2014/11806-9). MVCD thanks his scholarship from FAPESP (processes 2014/18632-6 and 2016/05254-9). CMdO and other Brazilian collaborators acknowledge funding from FAPESP through grants 2009/54202-8 and 2011/51680-6. WAS would like to acknowledge technical training fellowship number 2016/06484-8 from FAPESP and post-doc fellowship project 400247/2014-3 from the Brazilian funding agency CNPq. LSJr would like to acknowledge FAPESP for the fellowship number 2012/00800-4. J.A.H.J. thanks to Brazilian institution CNPq for financial support through  postdoctoral fellowship (project 150237/2017-0). EP acknowledges the Brazilian institution CNPq for the financial support through the technical training fellowship (project 155666/2016-9). ESC acknowledges financial support from Brazilian agencies CNPQ and FAPESP (process number 2014/13723-3). Y. J-T. acknowledges the financial support from the Funda\c{c}\~ao Carlos Chagas Filho de Amparo \`a Pesquisa do Estado do Rio de Janeiro - FAPERJ (fellowship Nota 10, PDR-10). RAD acknowledges support from CNPq through BP grant 312307/2015-2, CSIC through grant COOPB20263, FINEP grant REF 1217/13-01.13.0279.00 and REF 0859/10 - 01.10.0663.00 for partial hardware support for the J-PLUS project through the National Observatory of Brazil. We acknowledge the economical support from the Spanish Ministerio de Educaci\'on y Ciencia through grant AYA2006-14056 BES-2007-16280, AYA2012-30789, AYA2015-66211-C2-1-P and AYA2015-66211-C2-2. We thanks Dr. Cesar A. Caretta and Dr. Heinz Andermach for assisting us in the choice of target and for making the spectroscopy of the galaxies in the clusters available. This research has made use of the NASA/IPAC Extragalactic Database (NED) which is operated by the Jet Propulsion Laboratory, California Institute of Technology, under contract with the National Aeronautics and Space Administration. This work has made use of the computing facilities of the Laboratory of Astroinformatics (IAG/USP, NAT/Unicsul), whose purchase was made possible by the Brazilian agency FAPESP (grant 2009/54006-4) and the INCT-A. In particular, AM acknowledges Ulisses Manzo and Carlos E. Paladini at IAG/USP for helping with software installation. We thanks the anonymous referee for the careful reading of this manuscript and for her/his constructive comments and suggestions.
\end{acknowledgements}

\bibliographystyle{aa} 
\bibliography{ref}

\begin{appendix}
\section{Additional spectroscopic redshift information.}
\label{speczinfo}
\label{tableappendix}
This table indicates the different catalogs (all of them compiled by Andernach et al. 2005) that contributes to the A2589-A2593A dataset used in this paper, along with an estimate of the averaged limiting magnitude of each sample.

\vspace{0.5cm}

\begin{tabular}{|r|c|l}
\hline
  \multicolumn{1}{|c|}{Reference}  &
   \multicolumn{1}{|c|}{m$_{limit}$} \\
\hline
Humason   et al. 1956     & m$_{pg}$<13 \\
Hintzen   et al. 1980        &  m$_v$<16.0 \\
Barbon    et al. 1982        & m$_v$<17.0 \\
Huchra    et al. 1983        &  m$_v$<15.0 \\
Malumuth   \&  Kirshner 1985 &  - \\
Proust    et al. 1987        &  - \\
Bothun    et al. 1990        &   m$_v$<21.0 \\
Beers     et al. 1991        &   -  \\
Capelato  et al. 1991        &   m$_v$<16.0 \\
Huchra    et al. 1992        &  R<14.5 \\
Fouque    et al. 1993        &   V<15.0 \\
Giovanelli \&  Haynes 1993   & m$_z$<18.0\\
Fisher    et al. 1995        &  f$_{60}$>1.2 Jy\\
Owen      et al. 1995        &  R<23.0 \\
Postman   \&   Lauer 1995    & R$_c$<14.0 \\
Haynes    et al. 1997        &   - \\
Crawford  et al. 1999        &   - \\
Dale      et al. 1999        &   - \\
Dale      et al. 1999        &  - \\
Falco     et al. 1999        &   m$_{zw}$<15.5 \\
Huchra    et al. 1999        &   m$_b$<15.5 \\
Wegner    et al. 1999        &   R<16.0 \\
Saunders  et al. 2000        &   b$_j$<19.5  \\
Abazajian et al. 2004        &  R<18.0 \\
Smith     et al. 2004        &   R<17.0 \\
Tsvetkov  et al. 2004        &  - \\
Springob  et al. 2005        &   - \\
Theureau  et al. 2007        &   I$_{lim}$<13.0 \\
Loubser   et al. 2008        & m$_B$<15.5 \\
Aihara    et al. 2011        &  R<18 \\
Haynes    et al. 2011        & R<25.0 \\
Huchra    et al. 2012        & K$_s$<13.5 \\
\hline\end{tabular}

\end{appendix}

\end{document}